\newif\ifproofs
\newif\iffigures
\newif\ifonlineappendix
\onlineappendixtrue

\newif\ifnotfornow
\newif\ifkonektwo

\documentclass[12pt]{article}

\usepackage{amsfonts}
\usepackage{multicol}
\usepackage{amsthm}
\usepackage{graphics}
\usepackage{graphicx}
\usepackage{amsmath}
\usepackage{xcolor}
\usepackage{amssymb}
\usepackage{natbib}
\usepackage{bbm}
\usepackage{setspace}
\usepackage{tikz}
\usepackage[normalem]{ulem}
\usepackage{cancel}
\usepackage{pgfplots}
\pgfplotsset{compat=1.18}

\usetikzlibrary{arrows.meta,positioning,calc,fit,decorations.pathreplacing}

\usepackage{geometry}
\usepackage[colorlinks=true,linkcolor=blue,citecolor=blue,urlcolor=blue]{hyperref}

\definecolor{darkblue}{RGB}{0, 0, 139}
 \RequirePackage{hyperref}
 \hypersetup{
     colorlinks=true,
     linkcolor=darkblue,
     filecolor=darkblue,
     urlcolor=darkblue,
     citecolor=darkblue
 }

\title{Entry Deterrence with Public Signals:\\ Revisiting the Chain-Store Paradox}
\author{Francesc Dilm\'e and Aaron Kolb\footnote{Dilm\'e: University of Bonn, fdilme@uni-bonn.de. Kolb: Kelley School of Business, Indiana University, kolba@iu.edu. We thank Sarah Auster, Avi Lichtig, and Stephan Lauermann for their helpful comments, and the participants at the seminars at the University of Bonn, Indiana University, and the University of Wisconsin. This work was funded by a grant from the European Research Council (ERC 949465) and the Philpott Fellowship at Indiana University. Dilm\'e gratefully acknowledges the support from the German Research Foundation (DFG) through CRC TR 224 (Project B02) and under Germany’s Excellence Strategy – EXC 2126/1–390838866 and EXC 2047–390685813.}}
\date{November 3, 2026}

{\theoremstyle{plain}

\newtheorem{proposition}{Proposition}[section]
\newtheorem{lemma}{Lemma}[section]
\newtheorem{corollary}{Corollary}[section]
}

{\theoremstyle{remark}

\newtheorem{example}{Example}[section]
}

{\theoremstyle{definition}
\newtheorem{definition}{Definition}[section]
\newtheorem{assumption}{Assumption}

}

\newif\ifshieldagger
\shieldaggertrue

\ifshieldagger

\newcommand{\bankvariableso}[1]{#1^{\shield}}
\newcommand{\attackvariableso}[1]{#1^{\dagger}}

\else

\newcommand{\bankvariableso}[1]{{#1}^{\mathrm{o}}}
\newcommand{\attackvariableso}[1]{\tilde{#1}^{\mathrm{o}}}

\fi

\newcommand{\be}{\phi}

\newcommand{\bankhist}[1]{h^{#1}}
\newcommand{\cvar}{c}

\newcommand{\xvar}{x}

\newcommand{\bebank}{\bankvariableso{\be}}

\newcommand{\zbank}{\bankvariableso{z}}

\newcommand{\beattack}{\attackvariableso{\be}}

\newcommand{\zattack}{\attackvariableso{z}}

\newcommand{\gpay}{v}
\newcommand{\bebar}{\overline{\be}}

\newcommand{\beini}{\be_0}

\newcommand{\m}{\,}
\newcommand{\hatV}{V}
\newcommand{\hatW}{W}

\newcommand{\refrainact}{\mathrm{O}}
\newcommand{\attackact}{\mathrm{E}}
\newcommand{\giveupact}{\mathrm{A}}
\newcommand{\defendact}{\mathrm{F}}

\newcommand{\toughtype}{tough}
\newcommand{\atoughtype}{a tough}
\newcommand{\normaltype}{normal}

\newif\ifattackerdefender

\ifattackerdefender

\newcommand{\bank}{defender}
\newcommand{\Bank}{Defender}
\newcommand{\defendsv}{defends}
\newcommand{\defendv}{defend}
\newcommand{\defending}{defending}

\newcommand{\defense}{defense}
\newcommand{\defenses}{defenses}
\newcommand{\givesup}{concedes}
\newcommand{\giveup}{concede}
\newcommand{\givingup}{conceding}

\newcommand{\speculator}{attacker}

\newcommand{\speculators}{attackers}

\newcommand{\Speculator}{Attacker}

\newcommand{\attack}{attack}

\newcommand{\attackv}{attack}
\newcommand{\attacksv}{attacks}
\newcommand{\attacking}{attacking}

\newcommand{\refrain}{refrain}
\newcommand{\refrains}{refrains}
\newcommand{\refraining}{refraining}

\else

\newcommand{\bank}{monopolist}
\newcommand{\Bank}{Monopolist}
\newcommand{\defendsv}{fights}
\newcommand{\defendv}{fight}
\newcommand{\defending}{fighting}

\newcommand{\defense}{fight}
\newcommand{\defenses}{fights}

\newcommand{\givesup}{acquiesces}
\newcommand{\giveup}{acquiesce}
\newcommand{\givingup}{acquiescing}

\newcommand{\speculator}{entrant}

\newcommand{\speculators}{entrants}

\newcommand{\Speculator}{Entrant}

\newcommand{\attack}{entry}

\newcommand{\attackv}{enter}
\newcommand{\attacksv}{enters}
\newcommand{\attacking}{entering}

\newcommand{\refrain}{stay out}
\newcommand{\refrains}{stays out}
\newcommand{\refraining}{staying out}

\fi

\usepackage{tikz,pifont}
\usepackage{tikzsymbols}

\tikzsymbolsdefinesymbol {shield} { }
{%
\tikzsymbolsprovideandusesavebox{shield}{\begin{tikzpicture}[/tikzsymbolsstyle]
[baseline] \draw (.75ex,.83ex) -- (.4ex,1ex) -- (0.05ex,.83ex)  to [out=-90, in=150] (.4ex,0ex) to [out=30, in=-90] cycle;
\end{tikzpicture}%
}
}

\tikzsymbolsdefinesymbol {shieldp} { }
{%
\tikzsymbolsprovideandusesavebox{shieldp}{\begin{tikzpicture}[/tikzsymbolsstyle]
[baseline] \draw (.8ex,.83ex) -- (.4ex,1ex) -- (0ex,.83ex)  to [out=-90, in=150] (.4ex,0ex) to [out=30, in=-90] cycle;
\end{tikzpicture}%
}
}

\tikzsymbolsdefinesymbol {smallshield} { }
{%
\tikzsymbolsprovideandusesavebox{smallshield}{\begin{tikzpicture}[/tikzsymbolsstyle]
[baseline] \draw (.64ex,.9ex) -- (.4ex,1ex) -- (0.16ex,.9ex)  to [out=-90, in=150] (.4ex,0.36ex) to [out=30, in=-90] cycle;
\end{tikzpicture}%
}
}

\usepackage{enumitem}

\begin{document}

\maketitle
\begin{abstract}
We revisit the classic chain-store paradox by introducing exogenous public signals about the incumbent’s private type that arrive each period over a finite horizon. We fully characterize perfect Bayesian equilibria in terms of sequences of reputation cutoffs that increase over time: \speculators{} \attackv{} below their cutoff, while the normal incumbent mixes below its cutoff and \defendsv{} above it. The equilibrium outcome is generically unique. Information arrival dampens the incentive to build a reputation, so full deterrence need not emerge as the horizon grows arbitrarily long, even when the incumbent is very patient. We characterize exactly when there is full deterrence in the limit, along with the limiting equilibrium when there is not.
\end{abstract}

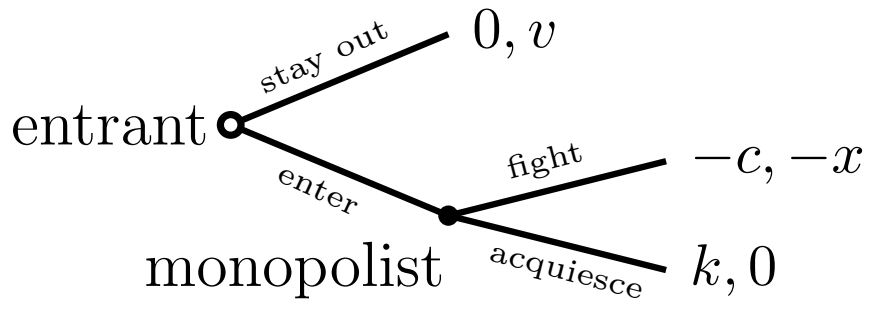
\begin{figure}
\begin{center}
    \begin{tikzpicture}[xscale=1.2,yscale=1.2]

    \draw[very thick] (1.25,1.15) node[right,rotate=24.6] {\tiny \refrain};
    \draw[very thick] (1.35,0.8) node[right,rotate=-24.6] {\tiny \attackv};
    \draw[very thick] (2.6,0.71) node[right,rotate=14.5] {\tiny \defendv};
    \draw[very thick] (2.5,0.33) node[right,rotate=-14.5] {\tiny \giveup};

    \draw[very thick] (1.2,1)--(2.4,1.5) node[right] {\small $0,\gpay$};
    \draw[very thick] (1.2,1)--(2.4,0.5);
    \draw[very thick] (2.4,0.5)--(3.6,0.8) node[right] {\small $-\cvar,-\xvar$};
    \draw[very thick] (2.4,0.5)--(3.6,0.2) node[right] {\small $k,0$};
   \draw [fill] (2.4,0.5) circle [radius=.05];
    \draw [fill] (2.5,0.2) node[left] {\bank};

  \draw [thick,fill] (1.2,1) circle [radius=.065];
  \draw [] (1.2,1) node[left] {\speculator};
  \draw [thick,fill,white] (1.2,1) circle [radius=.025];
\end{tikzpicture}
\end{center}
\caption{\citeauthor{selten1978chain}'s (\citeyear{selten1978chain}) chain-store game.}\label{figure_oneshot}
\end{figure}

\cite{selten1978chain} observed that deterrence games can exhibit implausible equilibrium behavior. He illustrated this with the chain-store game in Figure \ref{figure_oneshot}. In the one-period version of this game, an \speculator{} chooses to \refrain{} or \attackv{}. If it \attacksv{}, an incumbent \bank{} chooses to \defendv{} or \giveup. In the unique subgame-perfect Nash equilibrium of any finite repetition of this game---where the \bank{} faces a new \speculator{} in each period---all \speculators{} \attackv{} at all histories. \citeauthor{selten1978chain} argued this prediction was implausible, asserting that if \speculators{} observed multiple \defenses{} by the \bank{} in previous periods, they would adjust their behavior.

\cite{kreps1982reputation} and \cite{milgrom1982predation} addressed \citeauthor{selten1978chain}'s chain-store paradox by introducing a small uncertainty about the \bank's rationality. Their results are obtained in a model where the \bank{} is either \normaltype{}, with the payoffs specified in Figure \ref{figure_oneshot}, or \toughtype, in which case it always \defendsv{}. This alteration reverses \citeauthor{selten1978chain}'s prediction in long games: even with a small probability of \atoughtype{} \bank{}, there is full deterrence for an arbitrarily large number of initial periods as the time horizon lengthens. Intuitively, when the game's end is distant, the value from deterring future \attack{} is high, so the \normaltype{} \bank{} \defendsv{} current \attack{} to convince future \speculators{} that it is \toughtype. The anticipation of such behavior deters current \speculators{} even when the probability of the \bank{} being \toughtype{} is low.

We provide a new bridge between the extreme predictions from these two settings by taking the models of \cite{kreps1982reputation} and \cite{milgrom1982predation} and introducing a public signal about the \bank's type that arrives at the end of each period; the \cite{selten1978chain} model can be recovered in the limit as signals become perfectly revealing. We study perfect Bayesian equilibria and show that there is an essentially unique equilibrium outcome. We derive conditions under which full deterrence arises as the time horizon lengthens.

Our first result establishes that all equilibria feature cutoff strategies based on the posterior probability of the \bank{} being \toughtype{} (its reputation): There is a unique sequence of cutoffs $\beattack_t$ such that in any equilibrium, the period-$t$ \speculator{} \attacksv{} if the \bank's reputation is below  $\beattack_t$ and \refrains{} above $\beattack_t$. We show that deterrence is stronger, i.e., $\beattack_t$ is lower, in early periods of the game. We also show that the \bank{}'s continuation value is increasing in its reputation and in the number of periods remaining.

\phantomsection\label{intro:mechanism}While the value of reputation increases with the horizon length, the signal attenuates this growth. As shown in \cite{kreps1982reputation} and \cite{milgrom1982predation}, without a signal, the value of reputation increases significantly in the number of periods until the deadline, eventually leading to full deterrence---the Stackelberg outcome. The signal introduces a competing force: for a fixed horizon, the flow of information about the \bank{}'s type reduces the \bank{}'s incentive to build a reputation in the first place, because any reputation it builds is eroded by news. Indeed, by standard properties of learning, if the \bank{} is \normaltype{}, the posterior belief based on the signal process alone becomes arbitrarily close to zero with arbitrarily high probability at sufficiently late dates. Hence, the signal dampens the effect of a longer horizon.

We show that this second force need not cause deterrence to unravel. In fact, there can still be asymptotic full deterrence; that is, for any fixed $t$, $\beattack_t$ can converge to $0$ as the horizon lengthens to infinity. We show that this is the case if the signal is not too informative in a precise sense described below. Hence, despite the downward drift in the normal \bank{}'s reputation, its reputational incentives might remain sufficiently strong to deter \attack{} even when its reputation is initially low. However,  if the signal is sufficiently informative, there is no asymptotic full deterrence. We provide a sharp characterization in terms of the cost of \defending{}, $\xvar$ (as in Figure \ref{figure_oneshot}): there exists a threshold $\overline\xvar>0$ such that asymptotic full deterrence arises if and only if $\xvar\leq\overline\xvar$. Remarkably, the threshold $\overline\xvar$ converges to a finite limit as the discount factor $\delta$ approaches $1$. Thus, a sufficiently high fixed fighting cost precludes asymptotic full deterrence even for arbitrarily patient incumbents, despite the diverging payoff from permanent deterrence, $\gpay/(1-\delta)$.

We then connect the scope for asymptotic deterrence to the information structure of the public signals. We reinterpret the key deterrence threshold as a hitting-time problem for the log-likelihood ratio induced by news: under the normal type, the log-likelihood ratio drifts down as signals arrive, and there is asymptotic full deterrence if and only if the expected discounted time to cross a critical level is sufficiently long. This approach yields tractable sufficient conditions—stated in terms of the signal’s informativeness (via Kullback–Leibler divergence) together with an overshoot term—and it also delivers comparisons across signal structures in terms of their informativeness. We also show that while standard informativeness orders (such as Blackwell dominance) do not provide a general ranking of deterrence, our framework nests the familiar extremes: as signals become asymptotically revealing (resp. uninformative), long-run outcomes converge to the static benchmark (resp. the classic reputational full-deterrence limit).

\medskip

\noindent\textbf{Related literature.} Since the work of \cite{kreps1982reputation} and \cite{milgrom1982predation}, an extensive literature has been devoted to studying reputation in entry deterrence, reviewed in \cite{WILSON1992305}. More broadly, reputation building by long-lived players has been studied in numerous settings, most saliently in infinitely repeated games (see \citealp{mailath2006repeated} for a review).

Our paper examines implications of exogenous news arrival for deterrence. Earlier work has explored the role of exogenous news arrival in other signaling games. For example, \cite{daley2012waiting} study how exogenous news about an informed seller can cause delay of trade in equilibrium. Closer to our work, \cite{wiseman2009reputation} and \cite{hu2014reputation} study exogenous learning in infinite-horizon games with simultaneous moves and derive lower bounds on the long-lived player's payoff. These bounds do not, however, determine equilibrium behavior or payoffs, nor whether equilibrium outcomes of increasingly long finite-horizon games converge to full deterrence. We complement \citeauthor{wiseman2009reputation}'s (\citeyear{wiseman2009reputation}) and \citeauthor{hu2014reputation}'s (\citeyear{hu2014reputation}) payoff guarantees by studying the finite-horizon, sequential chain-store game: we characterize all perfect Bayesian equilibria, which share the same cutoff sequences and have a generically unique outcome, as well as their long-horizon limit, including exactly when asymptotic full deterrence occurs.\footnote{Other work studies the role of imperfect observation of \textit{actions} in reputation (e.g., \citealp{cripps2004imperfect}, \citealp{faingold2011reputation}, and \citealp{dilme2025repeated}).} Section \ref{subsec:wiseman_hu} provides a more detailed comparison with \cite{wiseman2009reputation} and \cite{hu2014reputation}.

Earlier work without exogenous news has studied other aspects of deterrence, including the role of endogenous types \citep{pitchik1993commitment} or nonconstant types (\citealp{aoyagi1996reputation} and \citealp{wiseman2008reputation}).

\section{Model}\label{section:model}

There is an incumbent \bank{} and a sequence of short-lived \speculators{}. In every period $t\in\{0,1,2,\dots,T\}$, the stage game depicted in Figure \ref{figure_oneshot} is played. The $t$-\speculator{} chooses to \attackv{} ($\attackact$) or \refrain{} ($\refrainact$). If the $t$-\speculator{} \attacksv, the \bank{} chooses whether to \defendv{} ($\defendact$) or \giveup{} ($\giveupact$).

The \bank{} has a private persistent type $\theta\in\{L,H\}$, where $H$ occurs with probability $\be_0\in(0,1)$. We model type $H$ as a behavioral ``\toughtype{}'' type that always \defendsv{} and type $L$ as a rational ``\normaltype{}'' type with payoffs defined below (we will often just use ``\bank{}'' to refer to the \normaltype{} \bank{}). At the end of each period, an exogenous public signal $s_t\in\mathbb R$ arrives, which is distributed i.i.d. conditional on $\theta$ and with distribution\vspace{-0.3cm}
\begin{align*}
  s_t\sim F_{\theta}(\cdot).
\end{align*}

In all periods $t$, the stage game payoffs are as follows. If the $t$-\speculator{} \refrains{}, it receives a payoff of 0 and the \bank{} obtains a benefit of $\gpay>0$. If the $t$-\speculator{} \attacksv{} and the \bank{} \defendsv{}, the $t$-\speculator{}'s payoff is $-\cvar<0$ and the \bank{}'s payoff is $-\xvar<0$. If the $t$-\speculator{} \attacksv{} and the \bank{} \givesup{}, the $t$-\speculator{}'s payoff is $k>0$ and the \bank{}'s payoff is $0$. The \bank{} maximizes the expected discounted sum of the stage game payoffs with discount factor $\delta\in(0,1)$. Throughout, we focus on the case where $\xvar<\frac{\delta}{1-\delta}\m\gpay$, as otherwise \defending{} is a strictly dominated strategy in any period for any $T$.

We impose the following assumption on the signal:
\begin{assumption}\label{assumption:signal}
The signal distributions have strictly positive densities $f_\theta(\cdot)$ on $\mathbb{R}$, and the log-likelihood ratio, $\log\frac{f_H}{f_L}$, is strictly increasing, has range $\mathbb R$, and has finite variance under both types.
\end{assumption}

This assumption ensures that, in each period and from any interior belief, the posterior belief based on the signal is continuously distributed with full support on $[0,1]$ under either type. Continuity ensures that any given interior belief, and in particular any cutoff belief, is reached with probability $0$ after the signal is realized.\footnote{We show that equilibria have a cutoff structure. A continuous posterior implies that there is zero probability that the belief at the beginning of the next period is exactly the \speculator{}'s cutoff, and therefore cutoffs in the current period do not depend on the way the \speculator{} breaks indifference in the next period.} Full support ensures that the expected continuation value from \defending{} is strictly increasing in the \bank{}'s reputation whenever the continuation value is increasing and nonconstant. We use this to establish uniqueness of the \bank{}'s cutoff.

\subsection{Strategies and equilibrium}

\noindent\textbf{Histories.} For $t\geq 0$, let $o_t\in\mathcal O:=\{\refrainact,(\attackact,\defendact),(\attackact,\giveupact)\}$ denote the outcome of period $t$ (not including the signal). For $t>0$, define $h^t=(o_0,s_0,\dots,o_{t-1},s_{t-1})$ as the history of all past actions and signals before period $t$, and set $h^0=\emptyset$. For $t>0$, define $H^t=(\mathcal O\times \mathbb{R})^t$, $H^0=\{\emptyset\}$, and $H=\cup_{t=0}^{T}H^t$.
\medskip

\noindent\textbf{Strategies.} An \emph{\speculator{}'s strategy} is a function $\alpha:H\to [0,1]$ mapping histories to \attack{} probabilities. Similarly, a \emph{\bank{}'s strategy} is a function $q:H\to [0,1]$ mapping histories to \defense{} probabilities conditional on \attack.\footnote{Since the \bank{} only moves in period $t$ if \attack{} occurs, we define $q$ only as a function of $h^t$, without explicitly appending ``$\attackact$'' to the history. Our equilibrium concept will require sequential rationality for the \bank{} after \attack{}.}
\medskip

\noindent\textbf{\Bank{}'s continuation payoffs.} Fix a strategy profile $(\alpha,q)$ and a history $h^t\in H$. The \emph{\bank's continuation value}, $V(h^t;\alpha,q)$, is defined recursively by
\begin{align*}
  V(h^T;\alpha,q)
  =
  (1-\alpha(h^T))\m
  \gpay
  +\alpha(h^T)\m q(\bankhist{T})\m(-\xvar)
\end{align*}
in the final period and\footnote{Because there is no risk of confusion and to simplify notation, we use $V(h^t,\attackact,d,s_t;\alpha,q)$ to denote $V((h^t,((\attackact,d),s_t));\alpha,q)$, where $(h^t,((\attackact,d),s_t))$ is the history entering period $t+1$ and $d\in \{\defendact,\giveupact\}$.}
\begin{align*}
  V(h^t;\alpha,q)
  =
  &(1-\alpha(h^t))\m
  (\gpay+\delta\m\mathbb{E}_L[V(h^t,\refrainact,s_t;\alpha,q)])\\
  &+\alpha(h^t)\m q(\bankhist{t})\m
  (\delta\m\mathbb{E}_L[V(h^t,\attackact,\defendact,s_t;\alpha,q)]
  -\xvar)\\
  &+\alpha(h^t)\m(1-q(\bankhist{t}))\m
  \delta\m\mathbb{E}_L[V(h^t,\attackact,\giveupact,s_t;\alpha,q)]
\end{align*}
for $t<T$, where expectations $\mathbb{E}_L$ are taken with respect to the signal $s_t$ under the type $L$. The terms on the right side correspond to the \bank's continuation payoff when the $t$-\speculator{} \refrains{} (first term), when the $t$-\speculator{} \attacksv{} and the \bank{} \defendsv{} (second term), and when the $t$-\speculator{} \attacksv{} and the \bank{} \givesup{} (third term).\medskip

\noindent\textbf{\Speculator's payoff.} The $t$-\speculator{}'s expected payoff from \attacking{} is $-\cvar$ if the \bank{} is \toughtype{} and
\begin{align*}
  q(\bankhist{t})\m(-\cvar)+(1-q(\bankhist{t}))\m k
\end{align*}
if the \bank{} is \normaltype. The payoff from \refraining{} is 0.

\subsubsection*{Perfect Bayesian equilibria}

Note that everything except for the \bank's type is public. Therefore, when defining perfect Bayesian equilibria, it is enough to focus on beliefs over the type, as each information set of each \speculator{} contains only two histories, one for each type.

\begin{definition}\label{def:PBE}
A \emph{perfect Bayesian equilibrium} consists of a strategy profile $(\alpha,q)$ and a belief system $\be:H\to [0,1]$ with $\be(\emptyset)=\beini$ such that, for all $h^t\in H$:
\begin{enumerate}
\item \Speculator{} optimality: $\alpha(h^t)=1$ and $\alpha(h^t)=0$ when
\begin{align*}
  \be(h^t)(-c)
  +(1-\be(h^t))\m\big(
  q(\bankhist{t})\m(-c)+
  (1-q(\bankhist{t}))\m k\big)
\end{align*}
is positive and negative, respectively.
\item \Bank{} optimality: $q(\bankhist{t})=1$ and $q(\bankhist{t})=0$ when
\begin{align*}
  \delta\m\mathbb{E}_L[V(h^t,\attackact,\defendact,s_t;\alpha,q)]-\xvar
  -\delta\m\mathbb{E}_L[V(h^t,\attackact,\giveupact,s_t;\alpha,q)]
\end{align*}
is positive and negative, respectively.

\item Bayes consistency: For all $(o_t,s_t)$ we have that $\be(h^t,o_t,s_t) =\hat\be_{s_t}(\be(h^t,o_t))$, where
\begin{align*}
  \hat\be_{s}(\be)
  :=\frac{\be\m f_H(s)}
  {\be\m f_H(s)
  +(1-\be)\m f_L(s)},
\end{align*}
where $\be(h^t,o_t)$ is obtained through Bayes' rule whenever possible, that is,
\begin{enumerate}
\item$\be(h^t,\refrainact)=\be(h^t)$,
\item$\be(h^{t},\attackact,\defendact)=\tfrac{\be(h^{t})}{\be(h^{t})+q(\bankhist{t})\m(1-\be(h^{t}))}$ if $\be(h^t)\neq 0$ or $q(\bankhist{t})\neq 0$,
\item $\be(h^{t},\attackact,\giveupact)=0$ if $\be(h^t)\neq 1$ and $q(\bankhist{t})\neq 1$.\footnote{In part 3, we abuse notation by extending the definition of $\be(\cdot)$ to histories that include the action(s) in period $t$ but not the signal.}
\end{enumerate}
\end{enumerate}
\end{definition}

Property 1 of Definition \ref{def:PBE} states that the $t$-\speculator{} \attacksv{} if the expected payoff from \attacking{} is positive, \refrains{} if negative, and can randomize otherwise. Property 2 says that the \bank{} \defendsv{} if the increase in discounted expected continuation value from \defending{} exceeds the cost of \defending{}, $\xvar$, and \givesup{} if it falls short. Property 3 says that beliefs update within a period first based on actions (resulting in $\be(h^t,o_t)$) and then based on the public signal (through $\hat\be_{s_t}$). If the \speculator{} \refrains{}, then beliefs update only from the signal. Bayes' rule always applies except in two situations: (i)  the \bank{} \givesup{} when $\be(h^t)=1$ or $q(\bankhist{t})=1$, and (ii) the \bank{} \defendsv{} when both $\be(h^t)=0$ and $q(\bankhist{t})=0$.

\section{Equilibrium analysis}

In this section, we characterize equilibrium behavior and show that it is essentially unique.

\subsection{Last period}\label{ss_lastperiod}
It is instructive to first analyze behavior in the last period. Fix an equilibrium $(\alpha,q,\be)$, and let $V(\cdot)$ denote $V(\cdot;\alpha,q)$.

To characterize the $t$-\speculator{}'s behavior, define\vspace{-0.15cm}
\begin{align}\label{eq_def_G}
    G(\hat\be,\hat q):=
    \hat\be\m(-c)
    +(1-\hat\be)(\hat q\m(-c)+(1-\hat q)\m k)
\end{align}
for all $\hat\be,\hat q\in[0,1]$; this is the $t$-\speculator{}'s expected payoff from \attacking{} when the \bank{}'s reputation is $\hat\be$ and the conjectured probability of \defending{} conditional on the \bank{} being \normaltype{} is $\hat q$. From the \speculator's optimality condition, the $t$-\speculator{} \attacksv{} for sure if $G(\be(h^t),q(h^t))>0$, and it \refrains{} for sure if $G(\be(h^t),q(h^t))<0$. Note that $G(\hat\be,\hat q)$ is  decreasing in $\hat\be$ and $\hat q$.

In the last period, the \bank{} \givesup{} after \attack, as \defending{} is costly and brings no future benefit. Thus, the \speculator{} \attacksv{} if $\be(h^T)<\bebar$ and \refrains{} if $\be(h^T)>\bebar$, where\vspace{-0.15cm}
\begin{align*}
    \bebar:=\frac{k}{c+k}
\end{align*}
is the unique belief such that $G(\bebar,0)=0$.

It follows that, in any equilibrium, the strategies at time $T$ are uniquely pinned down by the posterior $\be(h^T)$ except if $\be(h^T)=\bebar$. This implies that the \bank's continuation value satisfies\vspace{-0.15cm}
\begin{align}\label{eq_VT}
  V(h^T)
  \begin{cases}
    =0 & \text{if $\be(h^T)<\bebar$,}\\
    \in[0,\gpay] & \text{if $\be(h^T)=\bebar$,}\\
    =\gpay & \text{if $\be(h^T)>\bebar$}.\\
  \end{cases}
\end{align}

\subsection{Continuation at extreme beliefs}

We now provide a result that will significantly ease our analysis. It establishes that, if the \bank{} loses its reputation (i.e., $\be(h^t)=0$ at some history $h^t$) or if it perfectly convinces the \speculators{} about its type being \toughtype{} (i.e., $\be(h^t)=1$), then the continuation outcome coincides with the unique equilibrium outcome of the game in which the \bank{}'s type is common knowledge.

\begin{proposition}\label{prop:degen_beliefs_payoffs}
If the \bank{} has no reputation (i.e., $\be(h^t)=0$), its continuation payoff is 0 (i.e., $V(h^t)=0$): in all remaining periods, the \speculator{} \attacksv{} and the \bank{} \givesup. If the \bank{} is instead perceived to be \toughtype{} for sure (i.e., $\be(h^t)=1$), no \speculator{} \attacksv{} in the future and the \bank{}'s payoff is $V(h^t)=\frac{1-\delta^{T-t+1}}{1-\delta}\m\gpay$.
\end{proposition}

\subsection{Existence and uniqueness}
Strategies of the following cutoff form will play an important role in our analysis. As a mnemonic, we use the shield symbol {\Large $\shield$} for the \bank{} (the defender) and the dagger symbol $\dagger$ for the \speculator{} (the attacker).
\begin{definition}\label{def_eq_in_cutoffs}
An equilibrium $(\alpha,q,\be)$ is \emph{in cutoff strategies} with cutoffs $(\bebank_t,\beattack_t)_{t=0}^T$ if, for all $h^t\in H$, we have
\begin{align}\label{defender_cutoff_strategy}
  q(\bankhist{t})&=
    \begin{cases}
      1 & \text{if $\be(h^t)\geq\bebank_t$,}\\
     \frac{\be(h^t)(1-\bebank_{t})}{\bebank_t\m(1-\be(h^t))}
      & \text{if $\be(h^t)<\bebank_{t}$,}
    \end{cases}
\end{align}
whenever $\be(h^t)\neq 1$, and
\begin{align}\label{attacker_cutoff_strategy}
  \alpha(h^t)=
  \begin{cases}
      0 &  \text{if $\be(h^t)>\beattack_{t}$,}\\
      1 &  \text{if $\be(h^t)<\beattack_{t}$.}
  \end{cases}
\end{align}
\end{definition}

Note that Definition \ref{def_eq_in_cutoffs} does not impose any requirement on the \bank's strategy when $\be(h^t)=1$ or on the \speculators' strategy when $\be(h^t)=\beattack_t$. As we shall see, such flexibility will allow us to establish that all equilibria are in cutoff strategies. Note also that because no \speculator{} \attacksv{} if it believes that the \bank{} is \toughtype{} for sure, the value of $q(h^t)$ when $\be(h^t)=1$ is irrelevant for the equilibrium outcome. The value of $\alpha(h^t)$ when $\be(h^t)=\beattack_t$ will only be relevant at time 0 (in the non-generic case where $\be_0=\beattack_0$) because the probability that $\be(h^t)=\beattack_t$ for some $t>0$ will be 0 in all equilibria.

\begin{proposition}[Existence and uniqueness]\label{prop:uniqueness}
An equilibrium exists. There is a sequence $(\bebank_t,\beattack_t)_{t=0}^{T}$ (unique and independent of $\beini$) such that for all $\beini\in [0,1]$, all equilibria are in cutoff strategies with cutoffs $(\bebank_t,\beattack_t)_{t=0}^{T}$.
\ifproofs
\begin{proof}
See page \pageref{proof:prop:uniqueness}.
\end{proof}
\fi
\end{proposition}

While Proposition \ref{prop:uniqueness} establishes the uniqueness of the cutoffs of the equilibrium strategies, it does not, in general, guarantee that the continuation equilibrium outcome after each history is unique. The reason is that cutoff strategies do not specify the behavior of \speculators{} when they are indifferent (i.e., when $\be(h^t)=\beattack_t$).

Nonetheless, the equilibrium outcome (i.e., the ex ante distribution over terminal histories) is generically unique. To see this, observe that for all $t>0$, the probability that $\be(h^t)=\beattack_t$ is zero. This follows from the fact that $\beattack_t\in(0,1)$ for all $t$ and Assumption \ref{assumption:signal}.\footnote{Note that for any $h^{t}$ and $o_t$, either $\be(h^{t},o_t)\in\{0,1\}$ (and then the continuation outcome is uniquely pinned down by Proposition \ref{prop:degen_beliefs_payoffs}), or $\be(h^{t},o_t)\in(0,1)$ (and then $\be(h^{t},o_t,s_t)$ is continuously distributed on $[0,1]$ by Assumption \ref{assumption:signal}). That $\beattack_t\leq\bebar<1$ for all $t$ follows from equation \eqref{eq:beatt_bebar_beban}. That $\beattack_t>0$ for all $t$ is easily derived from the arguments in the proof of Proposition \ref{prop:uniqueness}.} This yields the following result.

\begin{corollary}
The equilibrium outcome is unique unless $\beattack_0=\beini$.\label{cor:outcome_unique}
\end{corollary}

\subsubsection*{Equilibrium construction}\phantomsection\label{par:eq_construction}
The proof of Proposition \ref{prop:uniqueness} works inductively from time $T$. In Section \ref{ss_lastperiod} we established that, in all equilibria, $q(h^T)$ and $\alpha(h^T)$ are of the form \eqref{defender_cutoff_strategy} and \eqref{attacker_cutoff_strategy} with cutoffs $\bebank_T=1$ and $\beattack_T=\bebar$. More generally, we show that whenever the \bank{}'s strategy has the form \eqref{defender_cutoff_strategy} for some $\bebank_t>0$, the total probability of \defending{} from the $t$-\speculator{}'s perspective at a belief $\be<\bebank_t$ is $\tfrac{\be}{\bebank_t}$; the $t$-\speculator{}'s indifference condition then implies its strategy has the form \eqref{attacker_cutoff_strategy} (Lemma \ref{lem:attacker_best_reply} in the Appendix) with
\begin{align}\label{eq:beatt_bebar_beban}
   \beattack_t=\bebar\m\bebank_t.
\end{align}

The induction argument then exploits this property by recursively showing that for each $t$ there is a unique function $\hatV_t:[0,1]\to[0,\frac1{1-\delta}\m\gpay]$ with the properties that (i) in any equilibrium, $V(h^t)=\hatV_t(\be(h^t))$ for all histories $h^t$ such that $\be(h^t)\neq\beattack_t$, and (ii)  $\hatV_t(\beattack_t)=0$.\footnote{It is convenient to choose $\hatV_t$ so that $\hatV_t(\be(h^t))$ coincides with the \bank's continuation value in an equilibrium where \speculators{} \attackv{} whenever they are indifferent to doing so (which always exists).} These functions are determined recursively by
\begin{align}\label{eq:hatV_cont}
  \hatV_t(\be)
  =\mathbb I_{\be>\beattack_t}
  \m(\gpay+\delta\m\mathbb E_L[
  \hatV_{t+1}(\hat\be_{s_t}(\be))]),
\end{align}
for $t<T$, where $\hatV_T$ is fully determined by the right side of \eqref{eq_VT} with $\hatV_T(\bebar)=0$ (as $\beattack_T=\bebar$). From Proposition \ref{prop:degen_beliefs_payoffs}, we have that $\hatV_t(0)=0$ and $\hatV_t(1)=\frac{1-\delta^{T-t+1}}{1-\delta}\m\gpay$ for all $t$.

If $\delta\m\hatV_{t+1}(1)\leq\xvar$, the \bank{} \givesup{} with certainty after \attack{} in period $t$ whenever $\be(h^t)<1$; in this case, the \bank{}'s strategy satisfies \eqref{defender_cutoff_strategy} with $\bebank_t=1$.\footnote{At belief one, \givingup{} is strictly optimal if $\delta\m\hatV_{t+1}(1)<\xvar$. If equality holds and the belief immediately after \givingup{} is zero, both actions yield zero, so either action or mixing is optimal. This does not affect the equilibrium outcome, since the \speculator{} \refrains{} at belief one.} This occurs, for example, in periods sufficiently close to the deadline. Otherwise, $\bebank_t$ is uniquely pinned down by
\begin{align}\label{eq:bank_indif}
  \delta\m\mathbb E_{L}
  [\hatV_{t+1}(\hat\be_{s}(\bebank_t))]=\xvar.
\end{align}
Indeed, by \eqref{defender_cutoff_strategy}, $q(h^t)\in(0,1)$ whenever $\be(h^t)\in (0,\bebank_t)$, so the \bank{} must be indifferent between \givingup{} and \defending{}, and by Bayes' rule, the posterior after a \defense{} is $\bebank_t$. Since by Proposition \ref{prop:degen_beliefs_payoffs} the \bank{} obtains 0 by \givingup, the continuation value after \defending{} (left side of \eqref{eq:bank_indif}) must be equal to the cost of \defending{} (right side of \eqref{eq:bank_indif}). This occurs, for example, in early periods if the deadline is long enough.\footnote{\label{fn:large_T_partial_separation}Because for any given $t$, $\delta\m\hatV_{t+1}(1)$ converges to $\frac{\delta}{1-\delta}\m\gpay>\xvar$ as $T\to\infty$, the condition that $\delta\m\hatV_{t+1}(1)>\xvar$ will hold if $T$ is high enough. Note that this condition and \eqref{eq:bank_indif} imply that $\bebank_0<1$ and $\beattack_0<\bebar$ for sufficiently large $T$.}

Based on this logic, equations \eqref{eq:beatt_bebar_beban}-\eqref{eq:bank_indif} fully determine the sequence $(\hatV_t,\beattack_t,\bebank_t)$, which is solved from $T$ backwards.

\subsection{Monotonicity of the continuation value}

The following proposition establishes monotonicity of the main equilibrium objects.

\begin{proposition}\label{prop:discrete_value_fn_monotone}
For all $t$, $\hatV_t(\be)$ is increasing in $\be$, and for all $\be\in(0,1)$, $\hatV_t(\be)$ is decreasing in $t$. Moreover, $\beattack_t$ and $\bebank_t$ are increasing in $t$.
\ifproofs
\begin{proof}
See page \pageref{proof:prop:discrete_value_fn_monotone}.
\end{proof}
\fi
\end{proposition}

The proposition first establishes that the \bank{}'s value function is increasing in its reputation level. This is intuitive: the \bank{}'s value equals the expected discounted per-period payoff $\gpay$ before the first entry, because its equilibrium continuation value conditional on entry is zero, and a higher reputation delays the first entry.

The proposition also establishes that the value function is monotone in $t$. Intuitively, near the deadline, there is less time remaining for the \bank{} to benefit from its reputation through a reduced probability of entry, so the continuation value weakly decreases at each belief. Hence, if the \bank{} still mixes at some beliefs, to preserve its indifference, it must induce a higher posterior belief when it \defendsv{}, as \eqref{eq:bank_indif} shows. Through \eqref{defender_cutoff_strategy}, this implies a lower probability of \defending{}. Thus, there is less deterrence near the deadline: by \eqref{eq:beatt_bebar_beban}, the \speculator{}'s cutoff increases over time.

\subsection{Example}\label{subsec:example}

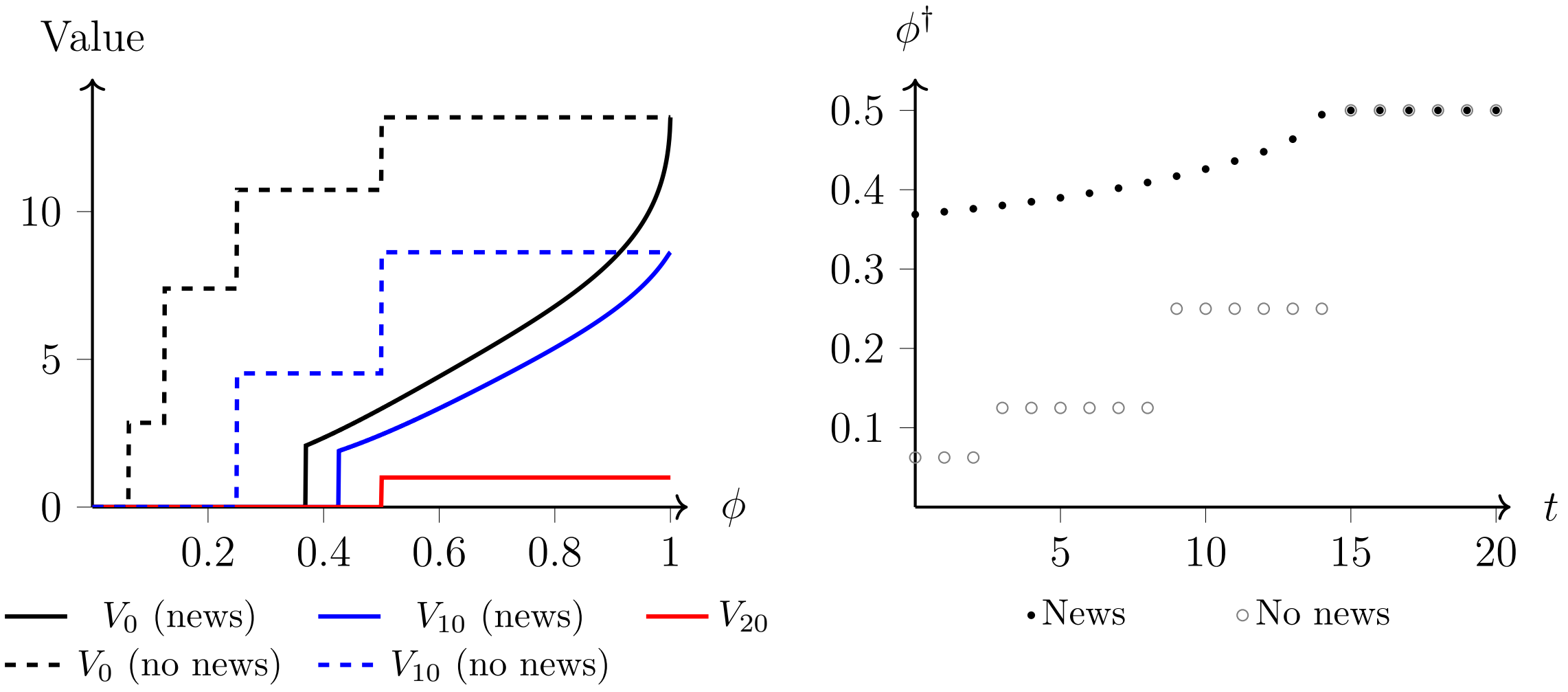
\begin{figure}[ht!]
\centering

\begin{minipage}[t]{0.48\linewidth}
\centering
\begin{tikzpicture}[baseline=(current axis.south)]
\begin{axis}[
    width=\linewidth,
    height=0.78\linewidth,
    axis lines=left,
    axis line style={->,thick},
    xmin=0,
    xmax=1.03,
    ymin=0,ymax=14.5,
    xtick={0.2,0.4,0.6,0.8,1},
    tick align=outside,
    tick style={black},
    xlabel={$\phi$},
    ylabel={Value},
    xlabel style={
        at={(axis description cs:1.03,0)},
        anchor=west
    },
    ylabel style={
        at={(axis description cs:0,1.03)},
        anchor=south,
        rotate=-90
    },
    grid=none,
   legend style={
    at={(0.5,-0.18)},
    anchor=north,
    draw=none,
    legend columns=3,
    font=\footnotesize,
    /tikz/every even column/.append style={column sep=0.25cm}
}
]

\addplot[black,very thick]
    table[x=phi,y=V0news] {value_data.dat};
\addlegendentry{$V_0$ (news)}

\addplot[blue,very thick]
    table[x=phi,y=V10news] {value_data.dat};
\addlegendentry{$V_{10}$ (news)}

\addplot[red,very thick]
    table[x=phi,y=V20] {value_data.dat};
\addlegendentry{$V_{20}$}

\addplot[black,very thick,dashed]
    table[x=phi,y=V0nonews] {value_data.dat};
\addlegendentry{$V_0$ (no news)}

\addplot[blue,very thick,dashed]
    table[x=phi,y=V10nonews] {value_data.dat};
\addlegendentry{$V_{10}$ (no news)}

\end{axis}
\end{tikzpicture}
\end{minipage}
\hfill
\begin{minipage}[t]{0.48\linewidth}
\centering
\begin{tikzpicture}[baseline=(current axis.south)]
\begin{axis}[
    width=\linewidth,
    height=0.78\linewidth,
    axis lines=left,
    axis line style={->,thick},
    xmin=0,
    xmax=20.5,
    ymin=0,
    ymax=0.54,
    xtick={5,10,15,20},
    ytick={0.1,0.2,0.3,0.4,0.5},
    tick align=outside,
    tick style={black},
    xlabel={$t$},
    ylabel={$\phi^\dagger$},
    xlabel style={
        at={(axis description cs:1.03,0)},
        anchor=west
    },
    ylabel style={
        at={(axis description cs:0,1.03)},
        anchor=south,
        rotate=-90
    },
    grid=none,
    legend style={
        at={(0.5,-0.18)},
        anchor=north,
        draw=none,
        legend columns=2,
        font=\footnotesize,
        /tikz/every even column/.append style={column sep=1cm}
    }
]

\addplot[
    only marks,
    mark=*,
    mark size=0.85pt,
    black
]
table[x=t,y=news] {cutoff_data.dat};
\addlegendentry{News}

\addplot[
    only marks,
    mark=o,
    mark size=1.5pt,
    gray
]
table[x=t,y=nonews] {cutoff_data.dat};
\addlegendentry{No news}

\end{axis}
\end{tikzpicture}
\end{minipage}

\caption{Equilibrium objects for $(k,c,\delta,\gpay,\xvar,\sigma,T)=(1,1,.95,1,5,1.5,20)$.} \label{fig:finite_T_plots}
\end{figure}

We now present an example of the equilibrium behavior for a particular signal structure. Identify $L$ with $0$ and $H$ with $1$, and suppose that each period's signal $s$ is normally distributed with mean $\theta$ and variance $\sigma^2$. Figure \ref{fig:finite_T_plots} shows the continuation value functions at $t\in \{0,10,20\}$ and the cutoff sequence $(\beattack_t)_{t=0}^T$ for a numerical example where the horizon is $T=20$. (We omit the \bank{}'s cutoff sequence because $\beattack_t=\bebar\bebank_t$, so plotting the \speculator{}'s cutoff conveys the same comparison.) For comparison, we show the corresponding value functions for the no-news benchmark model with dashed curves and the corresponding cutoffs with unfilled circles.\footnote{In the no-news benchmark, beliefs are constant until entry occurs, so jumps in the value function occur at beliefs where the first entry date is postponed. These cutoff beliefs can be reached with positive probability, so continuation values at the jump points can depend on history-dependent tie-breaking. The dashed curves show the values away from these discontinuities.\label{fn:no_news_jumps}} The figure illustrates that in this example, compared to the no-news benchmark, value functions are weakly lower and the \speculators{}' cutoffs are weakly higher. These comparisons are consistent with the dampening effect that news has on reputation-building incentives, particularly in early periods, when the number of future signals is large.

When signals are very noisy (e.g., $\sigma=30$), however, the value function with news can be higher at some beliefs and times than in the no-news benchmark. The reason is that a highly noisy signal process conveys little information about the \bank{}'s type but still introduces randomness in posterior beliefs.  This randomness can smooth out upward jumps in the continuation value. Thus, news can make the \bank{} better off slightly to the left of a jump and worse off slightly to the right. Similarly, the \speculator{}'s cutoff can be lower at some times with news than without.

\section{Long-horizon limit}

In this section, we present the main results of the paper. We examine the limit as $T\to \infty$ to understand whether long horizons still ensure deterrence in early periods, as they do in the model without exogenous news arrival.

Because equilibrium objects at time $t$ depend on $t$ and $T$ only through the number of remaining periods, $T-t$, an implication of Proposition \ref{prop:discrete_value_fn_monotone} is that $\beattack_{0}$ and $\bebank_{0}$ decrease as $T$ grows. Let $\beattack_*$ and $\bebank_*$ denote the limits of $\beattack_{0}$ and $\bebank_{0}$ as $T\to \infty$, respectively, and let $\hatV_*$ denote the pointwise limit of $\hatV_{0}$ as $T\to\infty$.\footnote{These limits exist because $\beattack_{0}$, $\bebank_{0}$, and $V_{0}$ are monotone in $T$ by Proposition \ref{prop:discrete_value_fn_monotone} and bounded.
} We say that there is \emph{asymptotic full deterrence} if $\beattack_*=0$. Instead, we say that there is \emph{asymptotic partial deterrence} if $\beattack_*>0$.\footnote{When there is asymptotic partial deterrence, it can be shown that as $T\to+\infty$, the expected fraction of periods with entry is $1$ under $L$.}

Note that the result in the models of \cite{kreps1982reputation} and \cite{milgrom1982predation}, where there is no exogenous signal, is that no \speculator{} \attacksv{} for an arbitrarily large number of initial periods as the time horizon lengthens for any prior belief; that is, there is asymptotic full deterrence.\footnote{Recall that we are assuming that $x<\frac{\delta}{1-\delta}\m\gpay$. Recall also that when $x\geq \frac{\delta}{1-\delta}\m\gpay$, \defending{} is a strictly dominated strategy regardless of the presence of a signal.} We will show that with a signal, there is asymptotic partial deterrence for some parameter values; in this case, if the prior belief on the type being \toughtype{} is low enough, the first \speculator{} \attacksv{} regardless of the length of the horizon.

\subsection{Limit equilibrium objects}

We now provide asymptotic analogues of the equilibrium conditions \eqref{eq:beatt_bebar_beban}-\eqref{eq:bank_indif} at $t=0$ (or any fixed $t$) by taking $T\to \infty$, and later show they characterize the limit equilibrium objects. These are given by the system of equations
\begin{align}
  \label{eq:stat_1}
  &\hatV(\be)
  =\mathbb I_{\be>\beattack}
  \m(\gpay+\delta\m\mathbb E_L[
  \hatV(\hat\be_{s}(\be))]),\\
  \label{eq:stat_2}
  &\beattack
  =\bebar\m\bebank,\\
  \label{eq:stat_3}
  &\delta\m\mathbb E_L
    [\hatV(\hat\be_{s}(\bebank))]
    =\xvar.
\end{align}
It is not difficult to see that equations \eqref{eq:stat_1}-\eqref{eq:stat_3} correspond to the equilibrium conditions of stationary equilibria in cutoff strategies with constant interior cutoffs $(\beattack,\bebank)$ in a model with infinite horizon. The following result relates solutions to equations \eqref{eq:stat_1}-\eqref{eq:stat_3} to the asymptotic equilibrium objects as $T\to\infty$.

\begin{proposition}\label{proposition:asymptotitc}
If \eqref{eq:stat_1}-\eqref{eq:stat_3} has a solution, it is unique and coincides with $(\hatV_*,\beattack_*,\bebank_*)$; moreover, $\beattack_*>0$. If, instead, there is no solution to \eqref{eq:stat_1}-\eqref{eq:stat_3}, then $\beattack_*=0$.
\end{proposition}

Proposition \ref{proposition:asymptotitc} provides a direct method for characterizing the long-horizon limits of finite-horizon equilibria: either there is no solution to \eqref{eq:stat_1}-\eqref{eq:stat_3}, in which case there is asymptotic full deterrence, or there is a unique solution, in which case there is asymptotic partial deterrence. Below, we show that these two possibilities are not vacuous.

\subsection{Conditions for asymptotic full deterrence}\label{subsection:conditions_asymptotic}

The following result establishes that asymptotic full deterrence is obtained if and only if the cost of \defending{} is at or below a threshold $\overline\xvar$.

\begin{proposition}\label{prop:conditions}
There is a threshold $\overline\xvar\in(0,\frac{\delta}{1-\delta}\m\gpay)$ such that there is asymptotic full deterrence if and only if $\xvar\leq\overline\xvar$.
\ifproofs
\begin{proof}
See page \pageref{proof:cor:conditions}.
\end{proof}
\fi
\end{proposition}

Proposition \ref{prop:conditions} is intuitive: asymptotic full deterrence occurs only if the cost of \defending{} is low enough. Importantly, the threshold cost $\overline\xvar$ is positive and smaller than the threshold $\frac{\delta}{1-\delta}\m\gpay$ (above which fighting is strictly dominated). That is, for any values of the remaining parameters, asymptotic full deterrence arises if the cost of \defending{} is low enough, and asymptotic partial deterrence arises if the cost of \defending{} is high enough.

Using the conditions \eqref{eq:stat_1}-\eqref{eq:stat_3} that apply in the asymptotic partial deterrence case, the threshold $\overline\xvar$ is obtained as follows. First, we change variables using the log-likelihood ratio $z\equiv\log(\be/(1-\be))$ instead of the belief $\be$. Let $\hatV(z;\zattack)$ denote the unique bounded solution to\footnote{Note that if $z(h^t)\in\mathbb R$ is the log-likelihood ratio at time $t$ and the $t$-\speculator{} does not \attackv, then the updated log-likelihood ratio after signal $s_t$ is $z(h^t)+\log(f_H(s_t)/f_L(s_t))$ by Bayes' rule.}
\begin{align}\label{e_fixed_point}
  \hatV(z;\zattack)
  =\mathbb I_{z>\zattack}
  \m\Big(\gpay+\delta\m\mathbb E_L\Big[
  \hatV\Big(z+\log \tfrac{f_H(s)}{f_L(s)};\zattack\Big)\Big]\Big),
\end{align}
which is \eqref{eq:stat_1} written in terms of $z$, with $\zattack=\log(\beattack/(1-\beattack))$. Changing variables to $\hat z\equiv z-\zattack$, we have $\hatV(z;\zattack)=U(z-\zattack)$, where $U(\hat z):=\hatV(\hat z;0)$ satisfies
\begin{align}\label{e_hatVstar0}
  U(\hat z)=\mathbb I_{\hat z>0}
  \m\Big(\gpay+\delta\m\mathbb E_L\Big[
  U\Big(\hat z+\log \tfrac{f_H(s)}{f_L(s)}\Big)\Big]\Big).
\end{align}
Note that $U$ does not depend on $\zattack$, and corresponds to the \bank{}'s payoff when $\hat z_0=\hat z$ and it obtains $\gpay$ in every period until the first period $\tau$ where $\hat z_\tau\leq0$ and zero thereafter (with $\tau=\infty$ if no crossing occurs). By standard arguments, $U$ is strictly increasing on $[0,\infty)$.

The indifference conditions for the \speculators{} and the \bank{} (\eqref{eq:stat_2} and \eqref{eq:stat_3}) imply that
\begin{align}\label{eq:VstarDelta}
  \xvar=\delta\m\mathbb E_L
  \Big[
  U\Big(\Delta(\beattack)
  +\log \tfrac{f_H(s)}{f_L(s)}\Big)\Big],
\end{align}
where, returning to using beliefs instead of log-likelihood ratios,
\begin{align}\label{eq:Delta_def}
\Delta(\beattack):=\log\bigg(\frac{\beattack/\bebar}{1-\beattack/\bebar}\bigg)
  -\log\bigg(\frac{\beattack}{1-\beattack}\bigg)
  =-\log\bigg(\frac{\bebar-\beattack}{1-\beattack}\bigg).
\end{align}

There exists a solution $(\hatV,\beattack,\bebank)$ to \eqref{eq:stat_1}-\eqref{eq:stat_3} with $\beattack>0$ if and only if there is a value $\beattack\in(0,\bebar)$ such that $\Delta(\beattack)$ satisfies equation \eqref{eq:VstarDelta}. Note that $\Delta(\cdot)$ is continuous and strictly increasing, tends to $+\infty$ as $\beattack\to\bebar$, and tends to $-\log(\bebar)>0$ as $\beattack\to0$. Hence, the right side of \eqref{eq:VstarDelta} is continuous and strictly increasing in $\beattack$ (because $U$ is nondecreasing and strictly increasing on $[0,\infty)$, and the log-likelihood ratio is atomless with full support), and it tends to $\delta\m\gpay/(1-\delta)>\xvar$ as $\beattack\to\bebar$. We then conclude that there is asymptotic partial deterrence if and only if $\xvar>\overline\xvar$, where
\begin{align}\label{eq:condition_fulldeterrence}
  \overline\xvar:=\delta\m\mathbb E_L\Big[
  U\Big(-\log\bebar
  +\log \tfrac{f_H(s)}{f_L(s)}\Big)\Big]
  =\delta\m\mathbb E_L\Big[
  \tfrac{1-\delta^{\tau-1}}{1-\delta}\m\gpay\Big],
\end{align}
where, under type $L$, $\tau$ is the first time $t$ at which $\hat z_t\leq0$, starting from $\hat z_0=-\log(\bebar)$. Since the log-likelihood ratio has full support on $\mathbb R$, we have $0<\mathbb P_L(\tau=1)<1$, so \eqref{eq:condition_fulldeterrence} implies $\overline\xvar\in(0,\frac{\delta}{1-\delta}\m\gpay)$.

Equation \eqref{eq:condition_fulldeterrence} is useful, as it immediately allows us to analyze the comparative statics of $\overline\xvar$ with respect to the parameters of the model. Intuitively, full deterrence is more likely to occur when the \bank{}'s value from deterrence is higher (higher $\delta$ or $\gpay$) or when it is easier to deter \speculators{} (higher $\cvar$ or lower $k$).
\begin{corollary}\label{cor:comp_statics_xbar}
The threshold $\overline\xvar$ is increasing in $\delta$, $\gpay$, and $\cvar$, and decreasing in $k$.
\end{corollary}

We explore the relationship between signal informativeness and deterrence in Sections \ref{subsec:wald} and \ref{subsec:informativeness}.

\subsubsection*{Patient \bank{} limit}

A notable implication of our analysis is that the threshold $\overline\xvar$ stays finite and strictly positive in the limit $\delta\uparrow 1$. That is, even in the limit as $\delta\uparrow 1$---which makes the value of permanent deterrence in the long-horizon limit, $\gpay/(1-\delta)$, grow arbitrarily large---there is a wide range of parameters for which the \bank{} cannot fully deter \attack. To see this, fix the signal structure and all parameters other than $\delta$ and $\xvar$. The first-passage representation \eqref{eq:condition_fulldeterrence} and monotone convergence give
\begin{align*}
  \lim_{\delta\uparrow1}\overline\xvar(\delta)
  =\gpay\m\bigl(\mathbb E_L[\tau]-1\bigr)<\infty,
\end{align*}
where finiteness of the expected crossing time is established below, and strict positivity follows because $\mathbb{P}_L(\tau>1)>0$ by Assumption \ref{assumption:signal}. Since $\overline\xvar$ is increasing in $\delta$ (Corollary \ref{cor:comp_statics_xbar}), fighting costs above this bound preclude asymptotic full deterrence even for arbitrarily patient incumbents.

\subsection{Information structures, hitting times, and deterrence}\label{subsec:wald}

In this section, we relate the threshold $\overline\xvar$ to the informativeness of the signal, and we establish sufficient conditions for one signal structure to be more conducive to full deterrence than another, in the sense of inducing a higher $\overline\xvar$.

Recall that $\overline\xvar$ can be obtained as follows. Let $(z_t)_{t=0}^\infty$ start from $z_0=-\log(\bebar)>0$, where each increment $z_{t+1}-z_t$ is independently and identically distributed according to
\begin{align}\label{eq_dist_z}
  z_{t+1}-z_t=\ell(s_{t+1}),
  \quad \text{with }s_{t+1}\sim F_L \text{ i.i.d.},
\end{align}
where $\ell(s):=\log\frac{f_H(s)}{f_L(s)}$ is the log-likelihood ratio. Then, $\overline\xvar$ is equal to the expected discounted payoff obtained from receiving $\gpay$ in periods $1,\ldots,\tau-1$; that is,
\begin{align*}
  \overline\xvar =\delta\m\mathbb E_L\bigg[
  \frac{1-\delta^{\tau-1}}{1-\delta}\bigg]\m\gpay, \quad \text{where} \quad \tau:=\inf\{t>0: z_t\leq 0\},
\end{align*}
which is the same as in \eqref{eq:condition_fulldeterrence}.

Unfortunately, $\mathbb E_L[\delta^\tau]$ is generally difficult to compute in closed form. Nevertheless, as $\delta\uparrow1$, the right side of equation \eqref{eq:condition_fulldeterrence} converges to $(\mathbb E_L[\tau]-1)\m\gpay$ by monotone convergence (recall that neither $-\log\bebar$ nor the distribution of $z_t-z_{t-1}$ depends on $\delta$, while finiteness of $\mathbb E_L[\tau]$ is verified below). We now use renewal theory to obtain an expression for the expectation of $\tau$. It is useful to define $O_\tau:=-z_\tau>0$, which is the distance by which the zero threshold is ``overshot'' at the first crossing time.

A key object in what follows is the log-likelihood process's drift under type $L$:
\begin{align*}
  \mathbb E_L[z_{t+1}-z_t]
  = \mathbb E_L[\ell(s)]
  = -\int f_L(s)\log\!\left(\frac{f_L(s)}{f_H(s)}\right)\m\mathrm ds.
\end{align*}
The integral is precisely the Kullback--Leibler (KL) divergence of $F_L$ with respect to $F_H$:
\begin{align*}
  D_{KL}(F_L\|F_H)
  :=\int f_L(s)\log\!\left(\frac{f_L(s)}{f_H(s)}\right)\m\mathrm ds,
\end{align*}
so that $\mathbb E_L[z_{t+1}-z_t] = -D_{KL}(F_L\|F_H)$. The KL divergence is a measure of the informativeness of the signal conditional on $\theta=L$, as it measures how different $F_L$ and $F_H$ are. A small value of $D_{KL}(F_L\|F_H)$ means that the signal is relatively uninformative in this sense. Note that by Jensen's inequality, we always have $D_{KL}(F_L\|F_H)\geq 0$, with strict inequality under Assumption \ref{assumption:signal}.

We can now use the Wald equation to express the expectation of $\tau$ as a function of the initial value $-\log\bebar$, the expected overshoot, and the KL divergence:
\begin{align}
  \mathbb E_L[\tau]=\frac{-\log\bebar+\mathbb E_L[O_\tau]}{D_{KL}(F_L\|F_H)}.
\label{eq:exp_crossing_time}
\end{align}

Intuitively, equation \eqref{eq:exp_crossing_time} says that the expected time it takes for $z$ to cross $0$ increases with the initial value $-\log\bebar$, while the effect of signal informativeness also depends on the expected overshoot. To derive equation \eqref{eq:exp_crossing_time}, first note that by the Wald equation (Theorem 2.6.2 in \citealp{durrett2019probability}), we have $\mathbb{E}_L[z_\tau-z_0]=\mathbb{E}_L[z_1-z_0]\m\mathbb{E}_L[\tau]$.\footnote{Specifically, define $X_i:=z_i-z_{i-1}$, $i=1,2,\dots$, and note that $X_1,X_2,\dots$ are i.i.d. By Assumption \ref{assumption:signal}, $\mathbb{E}_L[|X_1|]=\mathbb E_L[|\ell(s)|]$ is finite. Also, $\tau$ is a stopping time for $\mathcal F_n:=\sigma(X_1,\ldots,X_n)$. To verify $\mathbb E_L[\tau]<\infty$, note that $\mathbb E_L[e^{X_1}]=\int f_H(s)\,\mathrm ds=1$. Since $e^{X_1}$ is nonconstant, strict Jensen's inequality gives $\rho:=\mathbb E_L[e^{X_1/2}]<1$. For every $n\geq0$, the inclusion $\{\tau>n\}\subseteq\{z_n>0\}$, Markov's inequality, and independence imply
\[
  \mathbb P_L(\tau>n)
  \leq\mathbb E_L[e^{z_n/2}]
  =e^{z_0/2}\rho^n.
\]
Consequently,
\[
  \mathbb E_L[\tau]
  =\sum_{n=0}^\infty\mathbb P_L(\tau>n)
  \leq\frac{e^{z_0/2}}{1-\rho}<\infty.
\]
Hence, we can apply Theorem 2.6.2 in \cite{durrett2019probability}, which says that $\mathbb E_L[\sum_{t=1}^\tau X_t]=\mathbb E_L[X_1]\m\mathbb E_L[\tau]$.} Second, by the definitions of the log-likelihood ratio and KL divergence, $\mathbb{E}_L[z_1-z_0]=-D_{KL}(F_L\|F_H)$. Third, by the definition of the overshoot $O_\tau$, $\mathbb{E}_L[z_\tau-z_0]=\mathbb{E}_L[-z_0-O_\tau]=-(-\log\bebar+\mathbb{E}_L[O_\tau])$. The equation then follows from rearrangement.

We are now in a position to establish a condition that determines whether a signal induces asymptotic full or partial deterrence.

\begin{proposition}\label{prop_information_bound}
If
\begin{align}\label{eq_divergence_bound_rev}
  D_{KL}(F_L\|F_H)\geq\frac{\gpay}{\xvar+\gpay}\m
  (-\log\bebar+\mathbb E_L[O_\tau]),
\end{align}
then there is asymptotic partial deterrence. If \eqref{eq_divergence_bound_rev} fails and $\delta$ is sufficiently large, then there is asymptotic full deterrence.
\end{proposition}

An implication of Proposition \ref{prop_information_bound} is that if the signal informativeness (measured as $D_{KL}(F_L\|F_H)$) is below the threshold $-\frac{\gpay}{\xvar+\gpay}\m\log\bebar$ and $\delta$ is large enough, then there is asymptotic full deterrence. We then recover the classical result in \cite{kreps1982reputation} and \cite{milgrom1982predation} as the limit when the signal becomes uninformative: if $D_{KL}(F_L\|F_H)$ is small enough and the \bank{} is not too impatient, then for any fixed $\beini\in(0,1]$ and $T'\in\mathbb Z_+$, the probability that \attack{} occurs in the first $T'$ periods shrinks to $0$ in all equilibria as $T\to\infty$.

While equation \eqref{eq_divergence_bound_rev} provides a full characterization of asymptotic deterrence for large $\delta$, the term $\mathbb{E}_L[O_{\tau}]$ on the right side depends on both payoff parameters $\cvar$ and $k$ (through $\bebar=\frac{k}{c+k}$) and the signal structure. To simplify the comparison of deterrence across different signals, it is useful to establish a bound on $\mathbb{E}_L[O_{\tau}]$ that depends only on the signal structure. To this end, define
\begin{align*}
  \Delta_F:=-\inf_{a\geq0}\mathbb E_L
  \big[\ell(s)+a\big|\ell(s)+a\leq0\big]
  \in(0,\infty],
\end{align*}
which is the maximal mean residual life of $-\ell(s)$ over nonnegative thresholds and depends only on the signal structure. Assumption \ref{assumption:signal} does not guarantee that $\Delta_F$ is finite, even though $\mathbb E_L[O_\tau]$ is finite. Lemma \ref{lem:overshoot_bound} gives $0<\mathbb E_L[O_\tau]\leq\Delta_F$, and equality in the upper bound is possible. Then, a corollary of Proposition \ref{prop_information_bound} is the following.

\begin{corollary}\label{cor_information_bound}
If $D_{KL}(F_L\|F_H)\geq\frac{\gpay}{\xvar+\gpay}\m(-\log\bebar+\Delta_F)$, there is asymptotic partial deterrence.\footnote{This sufficient condition is vacuous when $\Delta_F=\infty$.} If $D_{KL}(F_L\|F_H)\leq-\frac{\gpay}{\xvar+\gpay}\m\log\bebar$ and $\delta$ is high enough, there is asymptotic full deterrence.
\end{corollary}

We finally provide a result comparing the scope for deterrence of two signals as a function of their informativeness.
\begin{proposition}\label{prop:KL_rank_xbar}
Assume $F$ and $G$ satisfy
\begin{align}\label{eq_prop_FG}
  \big(1+\tfrac{\Delta_F}{-\log\bebar}\big)^{-1}\m D_{KL}(F_L\|F_H)
  \geq
  D_{KL}(G_L\|G_H).
\end{align}
Then, if $\delta$ is large enough, $\overline\xvar$ is lower under $F$.\footnote{If $\Delta_F=\infty$, interpret the left side of \eqref{eq_prop_FG} as zero. The condition then cannot hold, since $D_{KL}(G_L\|G_H)>0$ under Assumption \ref{assumption:signal}.}
\end{proposition}

\paragraph{Relationship to \citeauthor{hu2014reputation}'s (\citeyear{hu2014reputation}) bounds.}
\cite{hu2014reputation} derives a lower bound on normalized equilibrium payoffs in an infinite-horizon game with simultaneous moves. The bound, specialized to chain-store payoffs and expressed in our notation, is
\[
\max\left\{0,\,v-\frac{v+x}{-\log\bebar}\left[D_{KL}(F_L\|F_H)+(1-\delta)\log\frac1{\beini}\right]\right\}.
\]
For informative signals, the limit of this bound as $\delta\to1$ is positive if and only if
\[
D_{\mathrm{KL}}(F_L\Vert F_H)
<
-\frac{\gpay}{\xvar+\gpay}\log\bebar.
\]
Moreover, even when this inequality holds, \citeauthor{hu2014reputation}'s (\citeyear{hu2014reputation}) bound converges to a value strictly below $\gpay$ as $\delta\uparrow 1$. Our Corollary \ref{cor_information_bound} gives a complementary conclusion: under this inequality, there is asymptotic full deterrence in our finite-horizon sequential game for sufficiently high $\delta$. Thus, for every fixed positive initial reputation, the normalized equilibrium payoff converges to $\gpay$ as $T\to\infty$. The same low-divergence restriction therefore yields full deterrence in our setting, rather than merely a guarantee of payoffs above minmax.

\begin{example}\label{example_KL_normal}
Consider a normal-normal experiment with variance $\sigma^2$ and mean $\mu_\theta$, with $\mu_H>\mu_L$. In this case, the log-likelihood ratio is
\begin{align*}
  \ell(s)=\log\frac{f_H(s)}{f_L(s)}
  =\frac{(\mu_H-\mu_L)\m(2\m s-\mu_H-\mu_L)}{2\m\sigma^2}.
\end{align*}
This implies that
\begin{align*}
  D_{\mathrm{KL}}(F_L\| F_H)
  =\frac{(\mu_H-\mu_L)^2}{2\m\sigma^2}.
\end{align*}
Furthermore,
\begin{align*}
    \Delta_F&=-\inf_{a\geq0}\mathbb E_L
\big[\ell(s)+a\big|\ell(s)+a\leq0\big]\\
  &=-\inf_{a\geq0}
  \bigg(a-D-\sqrt{2D}\m\frac{\varphi\big(-\frac{a-D}{\sqrt{2D}}\big)}{\Phi \big(-\frac{a-D}{\sqrt{2D}} \big)}\bigg)\\
  &=D+\sqrt{2D}\m\frac{\varphi\big(\sqrt{D/2}\big)}{\Phi\big(\sqrt{D/2}\big)},
\end{align*}
where $D:=D_{\mathrm{KL}}(F_L\| F_H)$ and $\varphi$ and $\Phi$ are the PDF and CDF of the standard normal distribution.\footnote{Note that $\log\frac{f_H(s)}{f_L(s)}$ is normally distributed under $L$ with mean $-D$ and variance $2D$.} The second equality follows from the standard formula for the mean of a truncated normal random variable. The third equality arises from the fact that the term inside the infimum is an increasing function of $a$ because $\frac{\mathrm{d}}{\mathrm{d}x}\frac{\varphi(x)}{\Phi(x)}\in(-1,0)$ for all $x\in\mathbb R$ (see \citealp{sampford1953some}).

We can therefore apply Corollary \ref{cor_information_bound} to the example in Section \ref{subsec:example}. For $(\mu_H,\mu_L,\sigma)=(1,0,1.5)$ and  $(k,c,\gpay,\xvar)=(1,1,1,5)$ as in Figure \ref{fig:finite_T_plots}, we have $D_{\mathrm{KL}}(F_L\| F_H)=\frac{2}{9}$, which is greater than $\frac{\gpay}{\xvar+\gpay}\m(-\log\bebar+\Delta_F)\approx 0.219$. Therefore, Corollary \ref{cor_information_bound} implies that there is asymptotic partial deterrence: the entrant's cutoff at time $0$ converges to a strictly positive limit as the horizon becomes arbitrarily long.
\end{example}

\subsection{Asymptotic informativeness and comparative statics}\label{subsec:informativeness}

In this section, we further explore how the informativeness of the signal affects deterrence and the \bank{}'s payoff for long horizons. We discuss  approaches using standard notions of informativeness, and then we present results for signal structures that become either very informative or very uninformative, approaching  the extreme cases of \cite{selten1978chain} on one hand and \cite{kreps1982reputation} and \cite{milgrom1982predation} on the other.

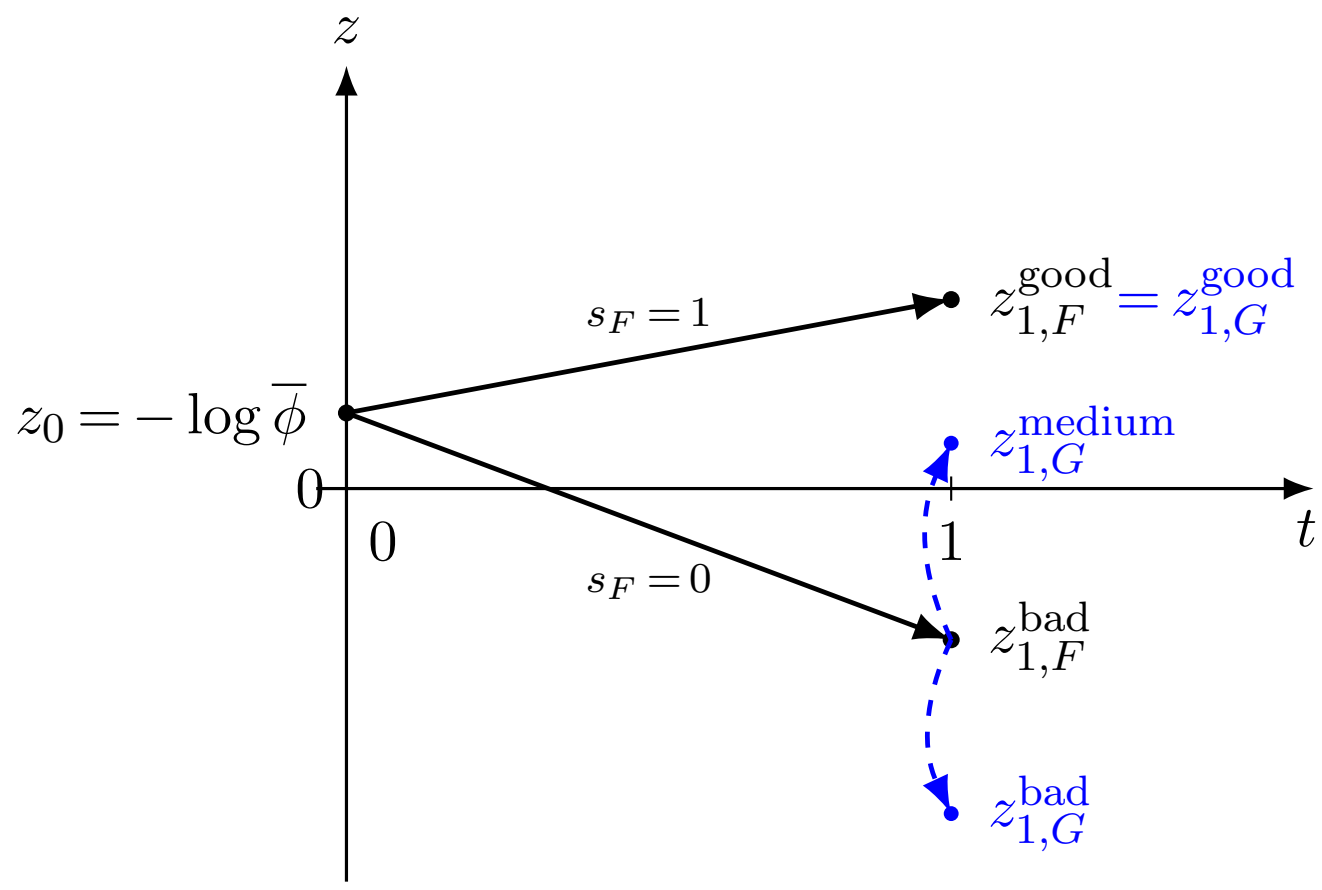
\begin{figure}[t]
\centering
\begin{tikzpicture}[
  font=\small,
  >=Latex,
  arrF/.style={-Latex, line width=0.9pt, black},
  split/.style={-Latex, line width=0.9pt, blue, dashed},
]

\draw[-Latex, line width=0.6pt] (-0.2,0) -- (6.4,0);
\draw[-Latex, line width=0.6pt] (0,-2.6) -- (0,2.8) node[above] {$z$};
\node[below] at (6.35,0) {$t$};

\draw[densely dotted] (0,0) -- (6.2,0);
\node[left] at (0,0) {$0$};

\draw (0,0.08) -- (0,-0.08) node[below right] {$0$};
\draw (4.0,0.08) -- (4.0,-0.08) node[below] {$1$};

\coordinate (z0)      at (0,0.5);
\coordinate (z1good)  at (4.0,1.25);
\coordinate (z1badF)  at (4.0,-1);

\coordinate (postHigh) at (4,0.3);
\coordinate (postLow)  at (4,-2.15);

\fill (z0) circle (1.6pt);
\node[anchor=east] at ($(z0)+(-0.12,0)$) {$z_0=-\log\bebar$};

\fill (z1good) circle (1.6pt)
  node[right=1mm] {$z_{1,F}^{\text{good}}{\color{blue}=z_{1,G}^{\text{good}}}$};

\fill (z1badF) circle (1.6pt)
  node[right=1mm] {$z_{1,F}^{\text{bad}}$};

\fill[blue] (postHigh) circle (1.4pt) node[right=1mm] {$z_{1,G}^{\text{medium}}$};
\fill[blue] (postLow)  circle (1.4pt) node[right=1mm] {$z_{1,G}^{\text{bad}}$};

\draw[arrF] (z0) -- (z1good) node[midway, above] {\scriptsize $s_F=1$};
\draw[arrF] (z0) -- (z1badF) node[midway, left=1mm,below=1mm] {\scriptsize $s_F=0$};

\draw[split] (z1badF) to[bend left=25] (postHigh);
\draw[split] (z1badF) to[bend right=25] (postLow);

\end{tikzpicture}
\caption{One-step log-likelihood update. Under $F$, $z_0$ moves to $z_{1,F}^{\text{good}}$ after $s_F=1$ or to $z_{1,F}^{\text{bad}}$ after $s_F=0$. A Blackwell-more-informative signal $G$ is constructed by splitting the $z_{1,F}^{\text{bad}}$ realizations into two posteriors: one above $0$ and another below $0$ (dashed blue arrows). Note that $F$ is a garbling of $G$, but $F$ induces a posterior log-likelihood ratio below $0$ with higher probability than does $G$.} \label{fig:garbling}
\end{figure}

To motivate our approach for studying informativeness below, let us comment on alternative approaches. First, Blackwell dominance is not strong enough to guarantee a lower $\overline\xvar$. We can see this by examining the last term in \eqref{eq:condition_fulldeterrence}, which is equal to the expected discounted time it takes for the log-likelihood ratio $z_t$ to first cross $0$ from $z_0=-\log(\bebar)>0$. The key is that a signal can be made Blackwell-more-informative by splitting the posterior belief after realizations $s$ satisfying  $\log\frac{f_H(s)}{f_L(s)}-\log(\bebar)<0$ to increase the probability that the posterior log-likelihood ratio lands above $0$, reducing the probability that an \attack{} occurs in period $1$. Then, if $\delta$ is small enough, such a modification makes the last term in \eqref{eq:condition_fulldeterrence} increase, thereby increasing the value of $\overline\xvar$ and expanding the set of parameters where asymptotic full deterrence occurs. In other words, making the signal more informative may change asymptotic partial deterrence into asymptotic full deterrence. Figure \ref{fig:garbling} illustrates an example using a binary signal for simplicity.

A second approach would be the following. For a fixed signal structure $f$, one would look for signal structures $g$ satisfying that the distribution of the likelihood ratio under $g$ conditional on $L$ first-order stochastically dominates the distribution of the likelihood ratio under $f$ conditional on $L$. It is easy to see that, in this case, the last term in \eqref{eq:condition_fulldeterrence} would be higher under $g$ than under $f$. Nevertheless, such a condition is vacuous: under Assumption \ref{assumption:signal}, the law of total probability ensures that only equality of the likelihood-ratio distributions under $L$ would satisfy this condition.\footnote{Indeed, note that for $h\in \{f,g\}$, $\int_{\mathbb R}\frac{h_H(s)}{h_L(s)}\m h_L(s)\m\mathrm{d}s=\int_{\mathbb R}h_H(s)\m\mathrm{d}s=1$, whereas the first-order stochastic dominance condition implies $\mathbb{E}^g_L[\frac{g_H(s)}{g_L(s)}]>\mathbb{E}^f_L[\frac{f_H(s)}{f_L(s)}]$ when their likelihood-ratio distributions differ.}

To summarize,  Blackwell dominance is too weak, while first-order stochastic dominance is too strong. To proceed, instead of defining a partial order over signal structures, we take an asymptotic approach. Given a signal structure with $s\sim f_{\theta}$ for $\theta \in\{H,L\}$, define the likelihood ratio $\Lambda_f(s):=\frac{f_H(s)}{f_L(s)}$. Consider   a sequence of signal structures with densities $f_{\theta}^n$ satisfying Assumption \ref{assumption:signal}, and let $\mathbb{P}^n_\theta$ denote the induced sequence of probability measures under type $\theta$. We say that $f_{\theta}^n$ is \emph{asymptotically revealing} if $\Lambda_{f^n}(s)\xrightarrow{\mathbb{P}^n_L} 0$, where $\Lambda_{f^n}(s)\xrightarrow{\mathbb{P}^n_L} a$ means that $\mathbb{P}^n_L(|\Lambda_{f^n}(s)-a|>\epsilon)\to0$ for every $\epsilon>0$. We say that it is \emph{asymptotically uninformative} if $\Lambda_{f^n}(s)\xrightarrow{\mathbb{P}^n_L} 1$.\footnote{For example, returning to the specification in Section \ref{subsec:example}, let $\sigma_n$ be a sequence of strictly positive real numbers and define $f^n_{\theta}$ to be the PDF of $N(\theta,\sigma_n^2)$. The sequence $f^n_{\theta}$ is asymptotically revealing if $\sigma_n\to 0$, and it is asymptotically uninformative if $\sigma_n\to \infty$. Note that for general $f^n_\theta$, the definition of asymptotically revealing implies that $\Lambda_{f^n}(s)\xrightarrow{\mathbb{P}^n_H} \infty$; likewise, the definition of asymptotically uninformative implies $\Lambda_{f^n}(s)\xrightarrow{\mathbb{P}^n_H} 1$.}  We use $\beattack{}^n_*$ and $V^n_*$ to denote the values of $\beattack_*$ and $V_*$ under $f^n_\theta$, respectively.

\begin{proposition}\label{prop:informativeness}
Let $f_{\theta}^n$ be a sequence of signal structures. If $f_{\theta}^n$ is asymptotically revealing, then: $\beattack{}^n_*\to \bebar$ and $\hatV^n_*(\be)\to\mathbb I_{\be\geq\bebar}\m\gpay$ for all $\be\in(0,1)$. If $f_{\theta}^n$ is asymptotically uninformative, then for large $n$, $\beattack{}^n_*=0$ and $\hatV^n_*(\be)=\mathbb I_{\be>0}\m\frac1{1-\delta}\m\gpay$ for all $\be\in[0,1]$.
\end{proposition}
An immediate implication of Proposition \ref{prop:informativeness} is that making the signals sufficiently informative (i.e., going sufficiently far along an asymptotically revealing sequence) hurts the \normaltype{} \bank{}, while making the signals sufficiently uninformative helps. Intuitively, when the \bank{} is \normaltype{} and signals become very informative, the \speculator{} learns quickly about the \bank{} from exogenous news. This makes a high reputation more transient for the \bank{}, reducing the \bank{}'s continuation value; this requires a lower probability of \defending{} to maintain indifference and therefore leads to \attack{} over a larger range of beliefs. In the limit, the \normaltype{} \bank's payoff converges to the payoff in the static game. In contrast, as signals become very uninformative, these forces act in the opposite direction, and the \bank{} is better off.

The proposition also implies that if the signal is not \textit{too} informative (i.e., it is sufficiently far along an asymptotically uninformative sequence), there is still asymptotic full deterrence. This follows from the convergence of the $t$-\speculator{}'s cutoff to $\beattack{}^n_*$ as $T\to \infty$, which is $0$ for sufficiently large $n$.

\section{Discussion}
\subsection{Welfare analysis}
To analyze the welfare effect of introducing exogenous news, consider the long-horizon limit, and assume $\xvar>\overline\xvar$, so that there is asymptotic partial deterrence with exogenous news, but asymptotic full deterrence without it. (Otherwise, there is asymptotic full deterrence both with and without exogenous news.) Assume that entrants' payoffs are aggregated and discounted by the same factor $\delta$ as the incumbent's. Since we do not specify consumer surplus, we first analyze producer surplus, i.e., the sum of the firms' payoffs.

It is immediate that exogenous news makes each entrant weakly better off in expectation, because entrants \refrain{} and obtain a payoff of $0$ in the no-news benchmark, and the option to \refrain{} remains available. Since entrants strictly prefer to enter under some histories that arise with positive probability with exogenous news, the total expected discounted payoff for the entrants is strictly positive with exogenous news. The same is true conditional on the incumbent being normal, since removing the tough type only increases the payoff from entry.

It is also immediate that exogenous news makes the normal incumbent strictly worse off in expectation, because it obtains the maximal possible payoff of $\gpay/(1-\delta)$ under full deterrence.

To compare these offsetting effects, we now analyze total expected producer surplus conditional on the incumbent being the normal type, which is the strategic type analyzed throughout the paper. Let $N_O$, $N_A$, and $N_F$ denote the expected discounted numbers of periods with no entry, accommodated entry, and entry followed by fighting, respectively. The equilibrium characterization determines these quantities and hence the incumbent's payoff $V$, the aggregate payoff of entrants $E$, and producer surplus $P$:
\[
V=\gpay N_O-\xvar N_F,\qquad
E=kN_A-\cvar N_F,\qquad
P=V+E.
\]
A concrete implication follows by comparing the long-horizon outcome with asymptotic full deterrence, under which producer surplus is $\gpay/(1-\delta)$. Since $N_O+N_A+N_F=1/(1-\delta)$ in the limit,
\[
P-\frac{\gpay}{1-\delta}
=
(k-\gpay)N_A-(\gpay+\xvar+\cvar)N_F.
\]
Introducing exogenous news that induces asymptotic partial deterrence has two effects. First, in each period with accommodated entry, the \bank{} forgoes $\gpay$ and the entrant earns $k$, for a net effect of $k-\gpay$, which is negative when $k<\gpay$. Second, in each period with fighting, the \bank{} forgoes $\gpay$ and both firms incur fighting costs, which unambiguously decreases producer surplus. Thus, whenever $\xvar>\overline \xvar$ and $k<\gpay$, the total producer surplus decreases; and since we have already argued that exogenous news raises entrants' expected discounted profits, the incumbent's losses more than offset these gains.

When $k>\gpay$, the sign of the change in producer surplus depends on the balance between accommodated entry and costly fighting, which the equilibrium characterization also determines.

Consumer surplus can be incorporated transparently, but requires additional assumptions. Let $u_O$, $u_A$, and $u_F$ denote per-period consumer surplus for the three possible stage-game outcomes; note that under full deterrence, discounted consumer surplus is $u_O/(1-\delta)$. Then the net effect on consumer surplus conditional on the normal type is
\begin{align*}
  u_O N_O+u_A N_A+u_F N_F-\frac{u_O}{1-\delta}=(u_A-u_O)N_A+(u_F-u_O)N_F.
\end{align*}
A sufficient condition for this to be nonnegative is that $u_A \geq u_O$ and $u_F\geq u_O$.\footnote{For example, in a richer model with demand and costs, $u_A\geq u_O$ could arise because entry increases competition or results in a more efficient producer, while $u_F\geq u_O$ could arise if fighting involves a price reduction.}

Total surplus $W$ changes from the benchmark by
\[
W-\frac{\gpay+u_O}{1-\delta}
=
[(k-\gpay)+(u_A-u_O)]N_A
+
[(u_F-u_O)-(\xvar+\cvar+\gpay)]N_F.
\]
If parameter values are such that this quantity is positive, total surplus increases; a simple sufficient condition is that the coefficients on $N_A$ and $N_F$, which are explicit functions of  parameters values, are positive. However, signing the change in welfare would require imposing more underlying structure on demand, costs, and the interpretation of fighting and accommodation.

\subsection{Relationship to \cite{wiseman2009reputation} and \cite{hu2014reputation}}\label{subsec:wiseman_hu}

\cite{wiseman2009reputation} studies an infinitely repeated, simultaneous-move chain-store game with symmetric binary signals. By bounding the frequency of entry along an always-fight deviation, he shows that sufficiently noisy signals guarantee a sufficiently patient normal incumbent a payoff above its minmax payoff in every sequential equilibrium. The bound approaches the monopoly payoff as signals become uninformative.

\cite{hu2014reputation} extends the analysis to general finite simultaneous-move stage games, private monitoring of actions, and multiple commitment types. Using the relative-entropy method of \cite{gossner2011simple}, he obtains an explicit payoff lower bound in every Nash equilibrium and for every discount factor. The bound depends on the prior, patience, and the KL divergence between the normal and commitment types' signal distributions. Both papers include public learning as a special case.

Our analysis concerns the same tension between reputation building through actions and reputation erosion through news, but characterizes behavior in the finite-horizon, sequential chain-store game with public signals about a fixed type. Whereas the earlier results provide payoff guarantees across equilibria, Proposition \ref{prop:uniqueness} establishes that all perfect Bayesian equilibria have the same cutoff sequences, and Corollary \ref{cor:outcome_unique} gives a generically unique outcome. Thus, our results determine not only payoffs but also the timing and probabilities of entry, fighting, and accommodation.

Our characterization also determines when asymptotic full deterrence fails. At each fixed discount factor, Proposition \ref{prop:conditions} shows that, as the horizon grows, the entry cutoff at any fixed date converges to zero if and only if $\xvar\leq\overline\xvar$, and to a strictly positive limit otherwise.\footnote{In particular, we show that for a fixed signal structure and fixed remaining payoff parameters, $\overline\xvar$ remains bounded as $\delta\uparrow1$. Thus, sufficiently high fixed fighting costs preclude asymptotic full deterrence even as the payoff from permanent deterrence, $\gpay/(1-\delta)$, becomes arbitrarily large.} The earlier payoff lower bounds do not determine when asymptotic full deterrence fails: a guarantee below the monopoly payoff remains compatible with full deterrence. Moreover, by Proposition \ref{proposition:asymptotitc}, equations \eqref{eq:stat_1}-\eqref{eq:stat_3} determine the limiting cutoffs and payoffs of finite-horizon equilibria.

The technical distinction is between bounding deviation payoffs and characterizing equilibrium incentives. \cite{hu2014reputation} uses relative entropy to obtain payoff guarantees from mimicking a commitment type, and his chain-store analysis uses log-odds dynamics to bound entry along an always-fight deviation. We instead use backward induction to determine equilibrium cutoffs and continuation values, and then characterize their long-horizon limits. The incumbent's indifference condition yields the exact first-passage representation of the critical fighting cost $\overline\xvar$ in \eqref{eq:condition_fulldeterrence}. Wald's equation then relates the patient-limit threshold to KL divergence and the expected overshoot. Thus, the additional contribution is an exact condition for asymptotic full deterrence, rather than the use of KL divergence or log-odds dynamics itself.

\subsection{Modeling choices}\label{subsec:modeling_choices}

Following \cite{selten1978chain}, \cite{kreps1982reputation}, and \cite{milgrom1982predation}, we study the finite-horizon, sequential chain-store game. This allows us to ask directly in the classical setting when public information prevents asymptotic full deterrence: when does the entry cutoff remain bounded away from zero across all equilibria as the horizon grows? Both the finite horizon and sequential timing matter for this question. Under the condition $\xvar<\delta\gpay/(1-\delta)$, the corresponding infinite-horizon sequential game admits a full-deterrence equilibrium regardless of the signal structure. With simultaneous moves, full deterrence can also be sustained by repeating a stage-game equilibrium, even with a finite horizon. The finite-horizon, sequential formulation instead allows us to fully characterize when public learning prevents full deterrence even as the deadline becomes arbitrarily distant.

The assumption that the exogenous signals about the \bank{} are public allows us to characterize equilibrium using the \bank{}'s public reputation and the remaining horizon as state variables. Consider instead signals whose entire history is observed by each \speculator{}, but not by the \bank{}. The \bank{} must then form beliefs about the \speculators{}' shared private information, updating these beliefs from observed entry and non-entry. As in the public-signal model, its fighting probability determines the \speculator{}'s entry cutoff. Now, however, this cutoff also determines what entry reveals about the signal history. Since future \speculators{} observe this history, the resulting inference affects the \bank{}'s continuation value from \defending{} after \attack{}, and hence its optimal fighting probability. This additional feedback introduces a within-period fixed-point problem: the entry cutoff determines the \bank{}'s fighting incentives through its beliefs, while the fighting probability must be consistent with that cutoff. We are pursuing this variant separately.

While we have considered a behavioral \toughtype{} type who \defendsv{} with certainty for simplicity, several extensions are of interest. A first possibility would be to consider a \toughtype{} type who \defendsv{} with a probability less than, but close to, $1$. This extension requires further analysis: \givingup{} would no longer reveal that the \bank{} is \normaltype{}, so an adjustment in the definition of $\bebar$ alone would not suffice. Another possibility would be to let the \toughtype{} type be a payoff type instead of a behavioral type. In line with \cite{kreps1982reputation}, this payoff type would prefer \defending{} to \givingup{} after \attack{} but would still prefer deterrence to \attack. In the Online Appendix, we show that, with such a payoff type, our equilibrium outcome remains an equilibrium outcome, and it is the unique one if the \toughtype{} type's payoff from \defending{} is sufficiently high.

\appendix

\section{Proofs}

\begin{proof}[Proof of Proposition \ref{prop:degen_beliefs_payoffs}]\label{proof:prop:degen_beliefs_payoffs}
Consider an arbitrary PBE $(\alpha,q,\be)$. We first show by induction that if $\be(h^t)=0$ for some $h^t\in H$, then (i) $V(h^t)=0$, (ii) $\alpha(h^t)=1$, and (iii) $q(\bankhist{t})=0$. Clearly, (i)-(iii) hold in the last period: we have $q(\bankhist{T})=0$, so if $\be(h^T)=0$, then $\alpha(h^T)=1$ and $V(h^T)=0$. Now suppose (i)-(iii) hold for all $t\geq \hat T+1$ for some $\hat T\geq 0$. We show that they also hold for $t=\hat T$. To rule out $q(\bankhist{\hat T})>0$, note that this would imply that $\be(h^{\hat T},\attackact,\defendact,s_{\hat T})=0$ for all $s_{\hat T}$ by Bayes' rule, and therefore $V(h^{\hat T},\attackact,\defendact,s_{\hat T};\alpha,q)=0$ by the induction hypothesis. But then \defending{} would yield $-\xvar<0$, whereas \givingup{} yields a nonnegative payoff (equilibrium continuation values are nonnegative, as the \bank{} can always \giveup{}), so the \bank{}'s optimality condition would imply $q(\bankhist{\hat T})=0$, a contradiction. Hence, $q(h^{\hat T})=0$, and so  $\alpha(h^{\hat T})=1$, so the \bank{}'s continuation payoff in period $\hat T$ is $0$ as desired.  This completes the induction argument. Moreover, note that if $\be(h^t)=0$, then since $q(h^t)=0$, Bayes' rule implies the \bank{}'s reputation stays at zero and (i)-(iii) apply for the remainder of the game.

Next, consider $\be(h^t)=1$. In this case, the \speculators{} \refrain{} every period and the posterior stays equal to 1. So, the \bank{} earns $\gpay$ in all periods $t,\dots,T$, for a discounted value of $V(h^t)=\frac{1-\delta^{T-t+1}}{1-\delta}\m\gpay$.
\end{proof}

\begin{lemma}\label{lem:attacker_best_reply}
Suppose that in period $t$, the \bank{}'s strategy has the form \eqref{defender_cutoff_strategy} for some $\bebank_t>0$. Then the $t$-\speculator{}'s best response has the form \eqref{attacker_cutoff_strategy}, where $\beattack_t=\bebar\m\bebank_t$.
\end{lemma}

\begin{proof}[Proof of Lemma \ref{lem:attacker_best_reply}]
\label{proof:lem:attacker_best_reply} Fix a history $h^t$ and a \bank{} strategy of the form \eqref{defender_cutoff_strategy} with fixed $\bebank_t>0$. If $\be(h^t)>\bebank_t$, the $t$-\speculator{}'s expected payoff from \attacking{} is $-c$. If instead $\be(h^t)\leq\bebank_t$, using \eqref{defender_cutoff_strategy} and simplifying yields
\begin{align*}
  G(\be(h^t),q(\bankhist{t}))
  =(\be(h^t)/\bebank_t)(-c)
   +(1-\be(h^t)/\bebank_t)k.
\end{align*}
When $\be(h^t)=\bebank_t=1$, this equality follows directly from $G(1,q)=-c$, irrespective of $q$. The displayed expression is strictly decreasing in $\be(h^t)$ on $[0,\bebank_t]$ and vanishes when $\be(h^t)=\frac{k}{c+k}\m\bebank_t=\bebar\m\bebank_t$. Thus, any best reply of the \speculator{} satisfies \eqref{attacker_cutoff_strategy} with $\beattack_t=\bebar\m\bebank_t$, with arbitrary mixing allowed at $\be(h^t)=\beattack_t$.
\end{proof}

\begin{proof}[Proof of Proposition \ref{prop:uniqueness}]\label{proof:prop:uniqueness}
We use an induction argument to derive the sequence of cutoffs referred to in the proposition and show that any equilibrium must be in cutoff strategies with these cutoffs. We then establish existence.\medskip

\noindent\textbf{Induction hypothesis for arbitrary $\hat T\leq T$:} For all $t\geq \hat T$, there is a unique threshold $\beattack_t\in(0,\bebar]$ and a function $\hatV_t:[0,1]\to\mathbb R_+$ such that in all equilibria, (i) the strategies at time $t$ satisfy equations \eqref{defender_cutoff_strategy}, \eqref{attacker_cutoff_strategy}, and \eqref{eq:beatt_bebar_beban}; (ii) $V(h^t)=\hatV_t(\be(h^t))$ whenever $\be(h^t)\neq\beattack_t$; and (iii) $\hatV_t(\beattack_t)=0$. Moreover, (iv) $\hatV_t$ is weakly increasing, $\hatV_t(\be)=0$ for all $\be\in[0,\beattack_t)$, and $\hatV_t(\be)>0$ for all $\be\in(\beattack_t,1]$.\medskip

\noindent\textbf{Base Case: $\hat T=T$.} The result holds for $\hat T=T$ as explained in Section \ref{ss_lastperiod}.\medskip

\noindent\textbf{Induction Step: Proof for $\hat T\in \{0,\dots,T-1\}$.} Assume that the induction hypothesis holds for $\hat T+1$. For all $t=\hat T,\ldots,T-1$, define
\begin{align*}
  \hatW_{t}(\be)
  :=\mathbb{E}_L[
  \delta\m\hatV_{t+1}(
  \hat \be_{s_t}(\be))]
\end{align*}
for all $\be\in[0,1]$ and $\hatW_{T}\equiv 0$. Because $\hatV_{t+1}$ is increasing and non-constant by the induction hypothesis and the signal distribution is continuous with a log-likelihood ratio that is strictly increasing and has full support on $\mathbb{R}$, we have that $\hatW_t$ is strictly increasing and continuous.

Note that for any history $h^{\hat T}$, the \bank{}'s expected utility from \defending{} is
\begin{align}
    -x+\hatW_{\hat T}(\be(h^{\hat T},\attackact,\defendact)). \label{eq:exp_payoff_defend}
\end{align}
There are two cases:
\begin{itemize}
\item \textbf{Case 1: $\hatW_{\hat T}(\be)<\xvar$ for all $\be\in[0,1]$.} In this case, for any history $h^{\hat T}$, \eqref{eq:exp_payoff_defend} is strictly negative, whereas the \bank{} can obtain at least $0$ by \givingup{}. Hence, the \bank{}'s optimality condition implies that $q(\bankhist{\hat T})=0$ for all $h^{\hat T}$. Then by the same logic as in Section \ref{ss_lastperiod}, it follows that $\alpha(h^{\hat T})=1$ if $\be(h^{\hat T})<\bebar$ and $\alpha(h^{\hat T})=0$ if $\be(h^{\hat T})>\bebar$. Hence, in this case, \eqref{defender_cutoff_strategy} and \eqref{attacker_cutoff_strategy} are satisfied at $t=\hat T$ with $\beattack_{\hat T}=\bebar$ and $\bebank_{\hat T}=1$. We define
\begin{align*}
    \hatV_{\hat T}(\be):=
    \begin{cases}
        0 &\text{if } \be\leq\bebar\\
        \gpay +\hatW_{\hat T}(\be) &\text{if }\be> \bebar.
    \end{cases}
\end{align*}

\item \textbf{Case 2: $\hatW_{\hat T}(\be)\geq\xvar$ for some $\be\in[0,1]$.} Because $\hatW_{\hat T}$ is continuous and strictly increasing, with $\hatW_{\hat T}(0)=0$, there is exactly one $\bebank_{\hat T}\in (0,1]$ for which $\hatW_{\hat T}(\bebank_{\hat T})=\xvar$.

We argue that the \bank{} plays according to \eqref{defender_cutoff_strategy} in period $\hat T$ with cutoff $\bebank_{\hat T}$, and we characterize belief updating. Assume $\be(h^{\hat T})<1$ (as \eqref{defender_cutoff_strategy} is only relevant in this case). First, if $\be(h^{\hat T})\geq \bebank_{\hat T}$, it must be that the \bank{} \defendsv{} with probability $1$, as in \eqref{defender_cutoff_strategy}. Toward a contradiction, assume the contrary. Bayes' rule would imply that $\be(h^{\hat T},\attackact,\giveupact)=0$, which, by Proposition \ref{prop:degen_beliefs_payoffs}, implies that the payoff from \givingup{} would be $0$.  Also, Bayes' rule would imply  $\be(h^{\hat T},\attackact,\defendact)> \be(h^{\hat T})\geq \bebank_{\hat T}$, so \eqref{eq:exp_payoff_defend} would be strictly positive.  Thus, \defending{} would be strictly optimal, a contradiction.

Second, if $0<\be(h^{\hat T})<\bebank_{\hat T}$, a similar contradiction arises if $\be(h^{\hat T},\attackact,\defendact)> \bebank_{\hat T}$; and if $\be(h^{\hat T},\attackact,\defendact)<\bebank_{\hat T}$, we would have $q(h^{\hat T})=0$, but then Bayes' rule would give $\be(h^{\hat T},\attackact,\defendact)=1\geq \bebank_{\hat T}$, again a contradiction. Hence, it must be that $\be(h^{\hat T},\attackact,\defendact)=\bebank_{\hat T}$, and Bayes' rule implies $q$ has the form in \eqref{defender_cutoff_strategy}. Third, if $\be(h^{\hat T})=0$, Proposition \ref{prop:degen_beliefs_payoffs} already establishes that $q(h^{\hat T})=0$, consistent with \eqref{defender_cutoff_strategy}.

Given $\bebank_{\hat T}$, by Lemma \ref{lem:attacker_best_reply}, the \speculator{} plays according to \eqref{attacker_cutoff_strategy} with $\beattack_{\hat T}=\bebar\m\bebank_{\hat T}$. In this case, define
\begin{align*}
    \hatV_{\hat T}(\be):=
    \begin{cases}
        0 &\text{if $\be\leq\beattack_{\hat T}$,}\\
       \gpay +\hatW_{\hat T}(\be) &\text{if $\be>\beattack_{\hat T}$}.
    \end{cases}
\end{align*}
\end{itemize}
We now establish properties (ii) and (iv) of the induction hypothesis for $\hat T$. Fix any history $h^{\hat T}$. If $\be(h^{\hat T})<\beattack_{\hat T}$, then \attack{} occurs for sure, and since $\beattack_{\hat T}<\bebank_{\hat T}$, the \bank{} is willing to \giveup{} as $q(h^{\hat T})<1$, so the \bank{}'s continuation payoff is equal to its payoff when it \givesup, which is $0$; thus $V(h^{\hat T})=0$ if $\be(h^{\hat T})<\beattack_{\hat T}$. Also, if $\be(h^{\hat T})>\beattack_{\hat T}$, we have
\begin{align*}
  V(h^{\hat T})
=v+\delta\m\mathbb{E}_L[V(h^{\hat T},\refrainact,s_{\hat T})]
  =v+\delta\m\mathbb{E}_L[\hatV_{{\hat T}+1}(\hat\be_{s_{\hat T}}(\be(h^{\hat T})))]~
  =v+\hatW_{\hat T}(\be(h^{\hat T})),
\end{align*}
where the second equality holds by the induction hypothesis for $\hat T+1$. In summary,
\begin{align*}
    V(h^{\hat T})=\begin{cases}
    0 & \text{if $\be(h^{\hat T})<\beattack_{\hat T}$,}\\
    \gpay+\hatW_{\hat T}(\be(h^{\hat T})) & \text{if $\be(h^{\hat T})>\beattack_{\hat T}$}.
  \end{cases}
\end{align*}
It is then clear that, for either Case 1 or Case 2 above, $\hatV_{\hat T}$ is the unique function satisfying  $V(h^{\hat T})=\hatV_{\hat T}(\be(h^{\hat T}))$ for all $h^{\hat T}\in H^{\hat T}$ such that $\be(h^{\hat T})\neq\beattack_{\hat T}$ and $\hatV_{\hat T}(\beattack_{\hat T})=0$. Moreover, $\hatV_{\hat T}$ is weakly increasing, and it satisfies $\hatV_{\hat T}(\be)=0$ for all $\be\in[0,\beattack_{\hat T}]$ and $\hatV_{\hat T}(\be)>0$ for all $\be\in(\beattack_{\hat T},1]$. This completes the induction argument.
\medskip

\noindent\textbf{Existence:} The previous argument establishes that equilibria are in cutoff strategies. Furthermore, the sequence of cutoffs $(\bebank_t,\beattack_t)_{t=0}^T$ pins down those strategies as a function of beliefs except at histories where $\be(h^t)=\beattack_t$ or $\be(h^t)=1$. To establish existence, we need only to specify the belief system and to specify the strategies and verify their optimality at $\be(h^t)=\beattack_t$ or $\be(h^t)=1$. Whenever $\be(h^t)=\beattack_t$, set $\alpha(h^t)=1$; since the \speculator{} is indifferent at its cutoff, this is optimal. When $\be(h^t)=1$, set $q(h^t)=0$ if $T-t< \tilde t$ and $q(h^t)=1$ if $T-t\geq \tilde t$, where $\tilde t> 0$ is defined as the unique solution to $\delta \frac{1-\delta^{\tilde t}}{1-\delta}\gpay=\xvar$. We define the belief system recursively going forward in time from $\be(\emptyset)=\be_0$. Given any $\be(h^t)$, set $\be(h^t,\attackact,\giveupact)=0$; note that this is implied by the Bayes' Consistency property when $\be(h^t)\neq 1$ and $q(h^t)\neq 1$. Set $\be(h^t,\attackact,\defendact)$ according to the Bayes' Consistency property when $q(h^t)\neq 0$ or $\be(h^t)\neq 0$, and set $\be(h^t,\attackact,\defendact)=0$ when $(\be(h^t),q(h^t))=(0,0)$. The \bank{}'s strategy is optimal by construction whenever $\be(h^t)\neq 1$. When $\be(h^t)=1$, note that under the given belief system, the \bank{}'s  payoff from \givingup{} is $0$, but if it \defendsv{}, it incurs the cost $\xvar$ and continues with a belief of $1$ in period $t+1$, for a discounted continuation payoff of $\delta\m\frac{1-\delta^{T-(t+1)+1}}{1-\delta}\m\gpay$ by Proposition \ref{prop:degen_beliefs_payoffs}. Hence, \defending{} is optimal if and only if $T-t\geq \tilde t$.
\end{proof}

\begin{proof}[Proof of Corollary \ref{cor:outcome_unique}]\label{proof:cor:outcome_unique}
The proof follows directly from Proposition \ref{prop:uniqueness} and the argument in the main text.
\end{proof}

\begin{proof}[Proof of Proposition \ref{prop:discrete_value_fn_monotone}]\label{proof:prop:discrete_value_fn_monotone}
In this proof, we will use the functions $\{\hatV_t,\hatW_t:[0,1]\to\mathbb R|t=0,...,T\}$ defined and used in the proof of Proposition \ref{prop:uniqueness}. Recall that all $\hatV_t$ satisfy that $\hatV_t(\be)=0$ for all $\be\leq\beattack_t$, $\hatV_t(\be)>0$ for all $\be>\beattack_t$, and $V_t$ is weakly increasing. Recall also that all $\hatW_t$ are strictly increasing for $t<T$ and continuous, and satisfy that if $\bebank_t\in(0,1)$, then $\hatW_t(\bebank_t)=\xvar$. We will use the following equality, which is valid for all $t\in\{0,...,T-1\}$ and $\be\in[0,1]$,
\begin{align*}
  \hatW_{t}(\be)
  =\mathbb E_L
  \big[\mathbb I_{\hat \be_{s_{t}}(\be)>\beattack_{t+1}}
  \m\delta\m(\gpay+\hatW_{t+1}(
  \hat \be_{s_{t}}(\be)))\big].
\end{align*}

We proceed by induction. \bigskip

\noindent\textbf{Induction hypothesis for $\hat T$:} For all $t\in\{\hat T,...,T-1\}$, we have (i) $\hatV_t(\be)\geq\hatV_{t+1}(\be)$ and $\hatW_t(\be)>\hatW_{t+1}(\be)$ for all $\be\in(0,1)$, (ii) $\bebank_{t}\leq\bebank_{t+1}$ and $\beattack_{t}\leq\beattack_{t+1}$.\bigskip

\noindent\textbf{Part 1.} We prove the result for $\hat T=T-1$. First, observe that $\bebank_{T-1}\leq1=\bebank_{T}$, which implies $\beattack_{T-1}=\bebar\m\bebank_{T-1}\leq\bebar=\beattack_{T}$. Also, because $\hatV_T(\be)$ is positive for all $\be>\bebar$, we have that $\hatW_{T-1}(\be)>0=\hatW_{T}(\be)$ for all $\be>0$. Finally, note that for $\be\leq \beattack_T$, $\hatV_{T}(\be)=0\leq \hatV_{T-1}(\be)$ and for $\be>\beattack_T\geq \beattack_{T-1}$, we have $\hatV_T(\be)=\gpay+\hatW_T(\be) \leq\gpay+\hatW_{T-1}(\be)=\hatV_{T-1}(\be)$.
\medskip

\noindent\textbf{Part 2.} Assume the induction hypothesis holds for $\hat T+1$ where $\hat T+1\leq T-1$. This implies that for all $\be\in(0,1)$
\begin{align*}
  \hatW_{\hat T}(\be)
  &=\mathbb E_L
  \big[\mathbb I_{\hat \be_{s_{\hat T}}(\be)>\beattack_{\hat T+1}}
  \m\delta (\gpay+\hatW_{\hat T+1}(
  \hat \be_{s_{\hat T}}(\be)))\big]\\
  &>\mathbb E_L
  \big[\mathbb I_{\hat \be_{s_{\hat T}}(\be)>\beattack_{\hat T+2}}
  \m \delta (\gpay+\hatW_{\hat T+2}(
  \hat \be_{s_{\hat T}}(\be)))\big]\\
  &=\hatW_{\hat T+1}(\be).
\end{align*}
The inequality holds for two reasons: $\beattack_{\hat T+2}\geq\beattack_{\hat T+1}$ (which shrinks the range where the integrand is non-zero) and, by assumption, $\hatW_{\hat T+2}(\hat \be_{s_{\hat T}}(\be))<\hatW_{\hat T+1}(\hat \be_{s_{\hat T}}(\be))$.

Since $\hatW_{\hat T}(\be)\geq\hatW_{\hat T+1}(\be)$ for all $\be\in[0,1]$ (for $\be\in\{0,1\}$, this follows from Proposition \ref{prop:degen_beliefs_payoffs}), we have that $\bebank_{\hat T+1}\geq\bebank_{\hat T}$ and so $\beattack_{\hat T+1}\geq\beattack_{\hat T}$.\footnote{Recall that, for all $t<T$, (i) if $\hatW_{t}(1)>\xvar$ then $\hatW_{t}(\bebank_{t})=\xvar$, (ii) if $\hatW_{t}(1)\leq\xvar$ then $\bebank_{t}=1$, and (iii) $\beattack_t=\bebar\m\bebank_t$.} It is also clear that $\hatV_{\hat T}(\be)\geq\hatV_{\hat T+1}(\be)$ for all $\be\in[0,1]$.
\end{proof}

\begin{proof}[Proof of Propositions \ref{proposition:asymptotitc} and \ref{prop:conditions}]\label{proof:proposition:asymptotitc} We divide the proof into parts.\medskip

\noindent\textbf{Part 0.} We first formalize several preliminary results from the discussion following Proposition \ref{prop:conditions}. First, we note that there is a unique bounded function $U$ satisfying equation \eqref{e_hatVstar0}.\footnote{For ease of exposition, we do not introduce equation \eqref{e_hatVstar0} until Section \ref{subsection:conditions_asymptotic}, but there is no circularity in our arguments.} Indeed, the functional
\begin{align*}
      \check V\mapsto
  \Big(\hat z\mapsto
  \mathbb I_{\hat z>0}
  \m\Big(\gpay+\delta\m\mathbb E_L\Big[
  \check V\Big(\hat z+\log\tfrac{f_H(s)}{f_L(s)}\Big)\Big]\Big)\Big)
\end{align*}
is a contraction in the space of functions from $\mathbb R$ to $[0,\frac1{1-\delta}\m\gpay]$ equipped with the sup norm. Note that $U(\hat z)$ is equal to the payoff obtained by the \bank{} if (i) the log-likelihood ratio of the initial belief is $\hat z$, (ii) the posterior evolves only according to the signal, and (iii) the \bank{} obtains $\gpay$ each period until the first period where $\hat z_t\leq0$, and obtains zero afterwards.

Also, for the reasons provided in the main text, there is some $\beattack\in(0,\bebar)$ satisfying equations \eqref{eq:VstarDelta} and \eqref{eq:Delta_def} if and only if $\xvar>\overline\xvar$, where $\overline\xvar$ is defined in equation \eqref{eq:condition_fulldeterrence}. Also, if $\xvar>\overline\xvar$, $\beattack$ is unique,  as the right side of \eqref{eq:VstarDelta} is strictly increasing in $\beattack$.\medskip

\noindent\textbf{Part 1.} We now consider the equilibrium behavior as $T$ grows. Let $(\beattack_{T,t})_{t=0}^T$  denote the \speculators' thresholds in the model with deadline $T$. Note that $\beattack_{T,t}=\beattack_{T-t,0}$ for all $T\in\mathbb Z_+$ and $t\leq T$. Note also that $\beattack_{T,t}$ is increasing in $t$. For a fixed $T$, letting $V_{T,0}$ denote the \bank{}'s value function at time $0$ for horizon $T$, note that
\begin{align*}
  \hatV_{T,0}(\be_0)
  =
  \frac{\mathbb E_L
  [1-\delta^{\tau_{T,0}}|\be_0]}
  {1-\delta}\m\gpay,
\end{align*}
where $\tau_{T,0}$ is the first time $t'$ such that $\be(h^{t'})\leq\beattack_{T,t'}$ (with $\tau_{T,0}\equiv T+1$ if no crossing occurs). Because $\beattack_{T,0}$ is decreasing in $T$, it converges to some $\beattack_{*}\geq0$ as $T\to\infty$.\medskip

\noindent\textbf{Part 2.} We analyze the case where $\beattack_*>0$. Let $(\be_t)$ represent beliefs based solely on the signals, using the same signal sequence for every horizon, and define $\tau_*:=\inf\{t\geq0:\be_t\leq\beattack_*\}$, with $\inf\varnothing=\infty$. Set $\tau_{T,0}=T+1$ if no cutoff is reached by time $T$.  Note that prior to the first entry time, $\be(h^t)=\be_t$. Since $\beattack_{T,t}=\beattack_{T-t,0}\downarrow\beattack_*$ for every fixed $t$, we have $\tau_{T,0}\uparrow\tau_*$ pathwise.\footnote{Indeed, $\tau_{T,0}\leq\tau_*$, and for every integer $n<\tau_*$, if $T$ is sufficiently large, all beliefs up to time $n$ exceed their corresponding finite-horizon cutoffs, implying that $\tau_{T,0}>n$.} Hence bounded convergence gives, for every $\be_0\in[0,1]$,
\begin{align*}
  \hatV_*(\be_0)
  =\lim_{T\to\infty}\hatV_{T,0}(\be_0)
  =\frac{\mathbb E_L[1-\delta^{\tau_*}\mid\be_0]}
         {1-\delta}\m\gpay,
\end{align*}
where $\delta^\infty:=0$. For $\be_0\in(0,1)$, this equals $U(z_0-\zattack_*)$ defined in \eqref{e_hatVstar0}, where $z_0=\log(\be_0/(1-\be_0))$ and $\zattack_*=\log(\beattack_*/(1-\beattack_*))$. At the endpoints, $\hatV_*(0)=0$ and $\hatV_*(1)=\gpay/(1-\delta)$.

By footnote \ref{fn:large_T_partial_separation}, $\bebank_{T,0}<1$ for all sufficiently large $T$, and the following indifference condition must hold:
\begin{align*}
  \xvar=\delta\m\mathbb E_L
  [\hatV_{T-1,0}(\hat\be_{s}(\bebank_{T,0}))].
\end{align*}
Taking $T\to \infty$ yields $\xvar=\delta\m\mathbb E_L[\hatV_{*}(\hat\be_{s}(\bebank_{*}))]$.\footnote{To justify the passage to the limit, define $y_T(s):=\hat\be_s(\bebank_{T,0})$ and $y(s):=\hat\be_s(\bebank_*)$. Since $\bebank_{T,0}\downarrow\bebank_*\in(0,1)$, we have $y_T(s)\downarrow y(s)$. Monotonicity in beliefs and horizons gives
\[
  \hatV_{T-1,0}(y(s))
  \leq\hatV_{T-1,0}(y_T(s))
  \leq\hatV_*(y_T(s)).
\]
The nondecreasing function $\hatV_*$ has at most countably many discontinuities, which the atomless law of $y(s)$ under Assumption \ref{assumption:signal} avoids almost surely. Both bounds therefore converge to $\hatV_*(y(s))$: the left by pointwise convergence and the right by continuity. Since the values are uniformly bounded by $\gpay/(1-\delta)$, dominated convergence applied to the indifference condition yields $\xvar=\delta\m\mathbb E_L\![\hatV_*(\hat\be_s(\bebank_*)) ]$.} Also, after a log-likelihood transformation, $\hat\be_{s}(\bebank_{*})$ becomes  $\zbank_{*}+\log \tfrac{f_H(s)}{f_L(s)}$, so $\hatV_{*}(\hat\be_{s}(\bebank_{*}))=U\big(\zbank_{*}+\log \tfrac{f_H(s)}{f_L(s)}-\zattack_{*}\big)$, which in turn equals $U\big(\Delta(\beattack_*) +\log \tfrac{f_H(s)}{f_L(s)}\big)$.

Putting these together, we have
\begin{align*}
  \xvar=
  \delta\m\mathbb E_L
  \Big[
  U\Big(\Delta(\beattack_*)
  +\log \tfrac{f_H(s)}{f_L(s)}\Big)\Big],
\end{align*}
which is \eqref{eq:VstarDelta} with $\beattack_*$ playing the role of $\beattack$. Because  $\beattack$ denotes the unique solution to \eqref{eq:VstarDelta}, we have $\beattack_*=\beattack$, and $(\hatV_*,\beattack_*,\bebank_*)$ is the unique solution to \eqref{eq:stat_1}-\eqref{eq:stat_3}. Also, recall that a solution with $\beattack>0$ exists if and only if $\xvar>\overline\xvar$.\medskip

\noindent\textbf{Part 3.} We now analyze the remaining case $\beattack_{*}=0$. Note that $\beattack=0$ cannot be part of a solution to \eqref{eq:stat_1}-\eqref{eq:stat_3}.\footnote{If there were a solution to this system with $\bebank=0$, then $\hat\be_{s}(\bebank)=0$. Since $\hatV(0)=0$ from \eqref{eq:stat_1}, the left side of \eqref{eq:stat_3} would also be $0$, which cannot equal $\xvar$.} Toward a contradiction, suppose there is a solution with $\beattack>0=\beattack_*$. Then, moving to log-likelihood space, there must be some $T$ such that $\zattack_{T,0}<\zattack\leq\zattack_{T-1,0}$. Since the finite-horizon cutoffs are increasing in time, $\zattack_{T-1,t}\geq\zattack_{T-1,0}\geq\zattack$ for $t=0,\ldots,T-1$. Using the same signal sequence in the two stopped-payoff representations therefore gives
\begin{align*}
  \hatV_{T-1,0}(z)\leq\hatV(z;\zattack)
  \qquad\text{for all }z\in\mathbb R,
\end{align*}
with strict inequality whenever $z>\zattack$.\footnote{For $z>\zattack$, there is strictly positive probability that the next $T$ log-likelihood increments are all positive. On this event, the constant-barrier process remains above $\zattack$ through period $T$ and receives $\gpay$ in that period, whereas the finite-horizon payoff ends at $T-1$. Since its stopped payoff also weakly dominates in all earlier periods, the expected discounted inequality is strict.} Moreover, $\bebank_{T,0}=\beattack_{T,0}/\bebar\in(0,1)$. Assumption \ref{assumption:signal} therefore implies
\[
  \mathbb P_L\!\left(
    \zbank_{T,0}+\log\tfrac{f_H(s)}{f_L(s)}
    >\zattack
  \right)>0.
\]
Thus,
\begin{align*}
  \xvar=\delta\m\mathbb E_L
  \Big[\hatV_{T-1,0}\Big(\zbank_{T,0}+\log \tfrac{f_H(s)}{f_L(s)}\Big)\Big]
  <\delta\m\mathbb E_L
  \Big[\hatV\Big(\zbank_{T,0}+\log \tfrac{f_H(s)}{f_L(s)};\zattack\Big)\Big].
\end{align*}
Because the right side of the inequality is increasing in $\zbank_{T,0}$ and is equal to $\xvar$ when $\zbank_{T,0}=\zbank$, we have $\zbank_{T,0}>\zbank$, which implies $\zattack_{T,0}>\zattack$. This contradicts that $\zattack_{T,0}<\zattack$.

We conclude that (when $\beattack_*=0$) there is no solution to \eqref{eq:stat_1}-\eqref{eq:stat_3}. Hence, it must be that $x\leq \overline x$. In this case, we have $\hatV_{T,0}(\be)\to \mathbb I_{\be>0}\m\frac1{1-\delta}\m\gpay$ pointwise as $T\to\infty$.\footnote{ Fix $\be>0$ and couple all horizons using the same signal sequence, with $(\be_t)$ evolving solely through signals from $\be_0=\be$. For every fixed integer $n\geq0$, Assumption \ref{assumption:signal} implies $\min_{0\leq t\leq n}\be_t>0$ almost surely, whereas $\max_{0\leq t\leq n}\beattack_{T,t}\to0$, since $\beattack_{T,t}=\beattack_{T-t,0}\to\beattack_*=0$. Thus $\tau_{T,0}\to\infty$ almost surely, and bounded convergence applied to the stopped-payoff representation in Part 1 gives $\hatV_{T,0}(\be)\to\gpay/(1-\delta)$. At $\be=0$, the value is zero for every horizon.}
\end{proof}

\begin{proof}[Proof of Corollary \ref{cor:comp_statics_xbar}]
Recall the last expression in \eqref{eq:condition_fulldeterrence}. The results with respect to $\gpay$ and $\delta$ are immediate. For $\cvar$ and $k$, couple the processes using the same sequence of log-likelihood increments. The hitting time $\tau$ is then pathwise nondecreasing in the initial state $-\log(\bebar)$, so \eqref{eq:condition_fulldeterrence} implies that $\overline\xvar$ is nondecreasing in this initial state. The inequality is strict: full support of the log-likelihood ratio gives positive probability that the process starting at the lower state stops at the first step while the other does not. Since $-\log(\bebar)$ is increasing in $\cvar$ and decreasing in $k$, the remaining results follow.
\end{proof}

\begin{proof}[Proof of Proposition \ref{prop_information_bound}]
Assume \eqref{eq_divergence_bound_rev} holds. Then
\begin{align*}
  \frac{\overline\xvar}{\gpay}=\delta\m\mathbb E_L\bigg[
  \frac{1-\delta^{\tau-1}}{1-\delta}\bigg]
  =\mathbb{E}_L\bigg[ \sum_{t=1}^{\tau-1}\delta^t\bigg]<\mathbb E_L[\tau]-1
  =\frac{-\log\bebar+\mathbb E_L[O_\tau]}{D_{KL}(F_L\|F_H)}
  -1
  \leq\frac{\xvar}\gpay,
\end{align*}
and by Proposition \ref{prop:conditions}, there is asymptotic partial deterrence.

If \eqref{eq_divergence_bound_rev} fails, we have
\begin{align*}
  \mathbb E_L[\tau]
  =\frac{-\log\bebar+\mathbb E_L[O_\tau]}{D_{KL}(F_L\|F_H)}
  >1+\frac{\xvar}\gpay.
\end{align*}
Also, as $\delta\uparrow 1$, by monotone convergence, we have \begin{align*} \frac{\overline\xvar}{\gpay}=\delta\m\mathbb E_L\bigg[ \frac{1-\delta^{\tau-1}}{1-\delta}\bigg] =\mathbb{E}_L\bigg[ \sum_{t=1}^{\tau-1}\delta^t\bigg]\to \mathbb{E}_L[\tau]-1>\frac{\xvar}{\gpay}.
\end{align*}

This implies that if $\delta$ is close enough to 1, then $\overline x>\xvar$, so there is asymptotic full deterrence.
\end{proof}

\begin{proof}[Proof of Corollary \ref{cor_information_bound}]\label{proof:cor_information_bound}
We begin by proving the following lemma that bounds the expected overshoot.
\begin{lemma}\label{lem:overshoot_bound}
We have $0<\mathbb{E}_L[O_\tau]\leq \Delta_F$.
\end{lemma}
\begin{proof}
The first inequality in the lemma follows from Assumption \ref{assumption:signal}. Toward the second inequality, since $\tau$ is the first time $z_t\leq 0$, we have $z_{\tau-1}>0$ and $z_\tau=z_{\tau-1}+X_\tau\leq 0$, where
\[
X_\tau:=z_\tau-z_{\tau-1}=\log\frac{f_H(s_\tau)}{f_L(s_\tau)}.
\]
We have $O_\tau=-z_\tau=-(z_{\tau-1}+X_\tau)$. For $a\geq0$, let $m(a):=\mathbb E_L[-(a+\ell(s))\mid a+\ell(s)\leq0]$, so that $\Delta_F=\sup_{a\geq0}m(a)$. Since $\{\tau=n\}=\{\tau>n-1\}\cap\{z_{n-1}+X_n\leq0\}$ and $X_n$ is independent of $(X_1,\ldots,X_{n-1})$, conditional on $\tau=n$ and $z_{n-1}=a$, the increment $X_n$ is distributed as $\ell(s)$ conditional on $a+\ell(s)\leq0$. Hence,
\begin{align*}
\mathbb E_L[O_\tau]
=\mathbb E_L\!\big[-(z_{\tau-1}+X_\tau)\big]
=\mathbb E_L\!\big[m(z_{\tau-1})\big]
\leq\sup_{a\geq0}m(a)
=\Delta_F.
\end{align*}
\end{proof}
\emph{(End of the proof of Lemma \ref{lem:overshoot_bound}. The proof of Corollary \ref{cor_information_bound} continues.)}

Assume first $D_{KL}(F_L\|F_H)\geq\frac{\gpay}{\xvar+\gpay}(-\log\bebar+\Delta_F)$. In this case, by Lemma \ref{lem:overshoot_bound},
\begin{align*}
    D_{KL}(F_L\|F_H)
    \geq\tfrac{\gpay}{\xvar+\gpay}(-\log\bebar+\Delta_F)\geq\tfrac{\gpay}{\xvar+\gpay}\m
    (-\log\bebar+\mathbb E_L[O_\tau]).
\end{align*}
Hence, by Proposition \ref{prop_information_bound}, there is asymptotic partial deterrence. Next, assume $D_{KL}(F_L\|F_H)\leq-\frac{\gpay}{\xvar+\gpay}\m\log\bebar$. In this case, again using Lemma \ref{lem:overshoot_bound},
\begin{align*}
    D_{KL}(F_L\|F_H)
    \leq-\tfrac{\gpay}{\xvar+\gpay}\m\log\bebar
    <\tfrac{\gpay}{\xvar+\gpay}\m
    (-\log\bebar+\mathbb E_L[O_\tau]).
\end{align*}
Hence, by Proposition \ref{prop_information_bound}, if $\delta$ is high enough, there is asymptotic full deterrence.
\end{proof}

\begin{proof}[Proof of Proposition \ref{prop:KL_rank_xbar}]
Assume \eqref{eq_prop_FG} holds. Then by Lemma \ref{lem:overshoot_bound},
\begin{align*}
  \mathbb E_L^F[\tau]
  =\frac{-\log\bebar+\mathbb E_L^F[O_\tau]}{D_{KL}(F_L\|F_H)}
  &\leq
  \frac{-\log\bebar+\Delta_F}{D_{KL}(F_L\|F_H)}\\
 & \leq\frac{-\log\bebar}{D_{KL}(G_L\|G_H)}\\
 &<\frac{-\log\bebar+\mathbb E_L^G[O_\tau]}{D_{KL}(G_L\|G_H)}
  =\mathbb E_L^G[\tau].
\end{align*}
When $\delta$ is close to 1, $\overline\xvar/\gpay+1$ under $F$ is close to the leftmost term and $\overline\xvar/\gpay+1$ under $G$ is close to the rightmost term. Thus, we have that $\overline x$ is lower under $F$.
\end{proof}

\begin{proof}[Proof of Proposition \ref{prop:informativeness}]\label{proof:prop:informativeness}

Fix a sequence of signal structures $f_{\theta}^n$. For each $n$, let $(\hatV_*^n,\beattack{}^n_*,\bebank{}^n_*)$ denote the limits of $(\hatV_0,\beattack_0,\bebank_0)$ as $T\to\infty$ under signal structure $f_{\theta}^n$. Write $\hatV^n(\cdot;\zattack)$ for the solution to \eqref{e_fixed_point} under this signal structure, and $U^n(\hat z)=\hatV^n(\hat z;0)$. Recall from Section \ref{subsection:conditions_asymptotic} that if $\beattack{}^n_*>0$, then
\begin{align}\label{eq:VstarDelta2}
  \xvar=\delta\m\mathbb E_L\big[
  U^n\big(\Delta(\beattack{}^n_*)
  +\log(\Lambda_{f^n}(s))\big)\big].
\end{align}

\noindent\textbf{Part 1.} Suppose the sequence $f_{\theta}^n$ is asymptotically revealing. Fix any $\hat{z}\in\mathbb R$ and $\epsilon\in(0,\min\{1,e^{-\hat z}\})$. By definition, there exists $N$ such that for $n\geq N$, $\Lambda_{f^n}(s)<\epsilon$, and hence $\hat{z}+\log(\Lambda_{f^n}(s))< 0$, with probability at least $1-\epsilon$ under $\theta=L$. Also, note that $U^n(\cdot)\in [0, \frac{\gpay}{1-\delta}]$ and $U^n(\hat z')=0$ for all $\hat z'\leq0$. Hence, for $n\geq N$, $\delta\m\mathbb E_L\big[U^n(\hat z+\log(\Lambda_{f^n}(s)))\big]\leq \epsilon\m\delta \frac{\gpay}{1-\delta}$. Therefore, $\delta\m\mathbb E_L\big[U^n(\hat z+\log(\Lambda_{f^n}(s)))\big]\to 0$ pointwise in $\hat z$ as $n\to \infty$.

This fact has three implications. First, by setting $\hat{z}=-\log(\bebar)$, we get that $\overline{x}^n$ in \eqref{eq:condition_fulldeterrence} tends to $0$ as $n\to\infty$. Hence, by Proposition \ref{prop:conditions}, for all sufficiently large $n$, we have $\beattack{}^n_*>0$; there is asymptotic partial deterrence. Second, and moreover, we must have $\beattack{}^n_*\to \bebar$. Toward a contradiction, if this were not true, then there would exist a subsequence of signal structures along which $\beattack{}^n_*$ is bounded away from $\bebar$ and therefore $\Delta(\beattack{}^n_*)$ is also bounded above by some $M>0$. But then setting $\hat{z}=M$ and taking $n$ sufficiently large, we have $\delta\m\mathbb E_L\big[U^n(\hat z+\log(\Lambda_{f^n}(s)))\big]<\xvar$, and since $U^n$ and $\Delta$ are nondecreasing, \eqref{eq:VstarDelta2} cannot be satisfied for any $\beattack{}^n_*$ along this subsequence. From this contradiction, we conclude that (along the original sequence) $\beattack{}^n_*\to \bebar$.

Third, for all fixed $\hat{z}$, we have $U^n(\hat z)\to \mathbb I_{\hat z>0} (\gpay+ 0)=\mathbb I_{\hat z>0} \m\gpay$ by \eqref{e_hatVstar0}. Using this fact, we now prove the claim that $\hatV^n_*(\be)\to\mathbb I_{\be\geq\bebar}\m\gpay$. Recall from Part 2 of the proof of Propositions \ref{proposition:asymptotitc} and \ref{prop:conditions} that, for all $n$ large enough that $\beattack{}^n_*>0$ and all $\be\in(0,1)$, $\hatV^n_*(\be)=U^n(z-\zattack{}^n_*)=\hatV^n(z;\zattack{}^n_*)$, where $z=\log(\be/(1-\be))$; let also $\overline z:=\log(\bebar/(1-\bebar))$. Fix any $z\geq \overline{z}$ and any $\epsilon>0$. For sufficiently large $n$, we have $\overline{z}-\epsilon<\zattack{}^n_*<\overline{z}\leq z$, where the first inequality follows from $\zattack{}^n_*\to \overline{z}$, and where $\zattack{}^n_*<\overline{z}$ follows from footnote \ref{fn:large_T_partial_separation}. Indeed, for each fixed $n$ with $\beattack{}^n_*>0$, footnote \ref{fn:large_T_partial_separation} gives a finite horizon $T_n$ such that $\zattack{}^n_{T_n,0}<\overline z$. Since $\zattack{}^n_{T,0}$ is nonincreasing in $T$,
\begin{align*}
  \zattack{}^n_*
  =\lim_{T\to\infty}\zattack{}^n_{T,0}
  \leq\zattack{}^n_{T_n,0}<\overline z.
\end{align*}
Thus, since $U^n$ is weakly increasing, $\hatV^n(z;\zattack{}^n_*)=U^n(z-\zattack{}^n_*)\leq U^n(z-(\overline{z}-\epsilon))\to \gpay$. Also, since $z\geq \overline z>\zattack{}^n_*,$ we have $\hatV^n(z;\zattack{}^n_*)\geq \gpay$ by \eqref{e_fixed_point} under signal structure $f_{\theta}^n$. Combining these facts gives  $\hatV^n(z;\zattack{}^n_*)\to \gpay$. And for any $z<\overline{z}$ and $\epsilon<\overline{z}-z$, for sufficiently large $n$, $\zattack{}^n_*>\overline{z}-\epsilon>z$. Thus, for $z<\overline z$ and sufficiently large $n$, $\hatV^n(z;\zattack{}^n_*)=U^n(z-\zattack{}^n_*)= 0$.

\medskip
\noindent\textbf{Part 2.} Next, suppose the sequence $f_{\theta}^n$ is asymptotically uninformative. Let $\mathbb{E}^n_L$ denote expectations under the low type and $\overline\xvar^n$ the value of $\overline\xvar$ for a given signal structure $f_{\theta}^n$. Recall that
\begin{align*}
\overline\xvar^n=\delta\m\mathbb E^n_L\Big[
  \frac{1-\delta^{\tau-1}}{1-\delta}\m\gpay\Big],
  \end{align*}
where $\tau$ is the first time $t>0$ at which the process $(z_t)$ in \eqref{eq_dist_z}, started from $z_0=-\log\bebar>0$, satisfies $z_t\leq0$. We show that $\overline\xvar^n\to \frac{\delta\m\gpay}{1-\delta}$. First, we claim that $\tau\xrightarrow{\mathbb{P}^n_L}\infty$. Let $X_t=z_t-z_{t-1}$. Note that $X_1,X_2,\dots$ are i.i.d. under $\mathbb{P}^n_L$ for each $n$. Then for any $T>0$, we have
\begin{align}
    \mathbb{P}^n_L(\tau\leq T)\leq T \,\mathbb{P}^n_L(X_1\leq -z_0/T)\to 0,
\end{align}
where the final convergence uses that $\Lambda_{f^n}(s)\xrightarrow{\mathbb{P}^n_L}1$, so $X_1=\log\Lambda_{f^n}(s)\xrightarrow{\mathbb{P}^n_L}0$. In other words, $\tau$ converges to $+\infty$ in $\mathbb{P}^n_L$-probability. Since $\delta^{\tau-1}\in [0,1]$, it follows that $\mathbb{E}^n_L[\delta^{\tau-1}]\to 0$, so $\overline\xvar^n\to \frac{\delta\m\gpay}{1-\delta}>\xvar.$

By Proposition \ref{prop:conditions}, there exists $N$ such that $\beattack{}^n_*=0$ for all $n\geq N$. For each such $n$, the full-deterrence value limit established in Part 3 of the proof of Propositions \ref{proposition:asymptotitc} and \ref{prop:conditions} gives
\begin{align*}
  \hatV^n_*(\be)
  =\lim_{T\to\infty}\hatV^n_{T,0}(\be)
  =\mathbb I_{\be>0}\m\frac{\gpay}{1-\delta},
  \qquad \be\in[0,1].
\end{align*}
\end{proof}

\end{document}